\documentclass[journal,12pt,draftclsnofoot,onecolumn]{IEEEtran}

\usepackage{amsfonts}
\usepackage{amsmath}
\usepackage{amssymb}
\usepackage{cite}
\usepackage{dsfont}
\usepackage{latexsym}
\usepackage{subfigure}
\usepackage{times}
\usepackage{url}
\usepackage{verbatim}

\usepackage{graphicx}

\def\bb0{{\mathbb{0}}}

\def\bb{{\mathbf{b}}}

\def\bi{{\mathbf{i}}}

\def\bp{{\mathbf{p}}}

\def\br{{\mathbf{r}}}
\def\bs{{\mathbf{s}}}

\def\by{{\mathbf{y}}}

\def\b0{{\mathbf{0}}}

\def\bA{{\mathbf{A}}}

\def\bD{{\mathbf{D}}}
\def\bE{{\mathbf{E}}}
\def\bF{{\mathbf{F}}}

\def\bI{{\mathbf{I}}}

\def\bR{{\mathbf{R}}}

\def\sf0{{\mathsf{0}}}

\def\rmD{\mathrm{D}}

\def\rmT{\mathrm{T}}
\def\rmU{\mathrm{U}}
\def\rmV{\mathrm{V}}

\def\rmc{{\mathrm{c}}}

\def\rmn{{\mathrm{n}}}

\def\rmq{{\mathrm{q}}}

\def\rmt{{\mathrm{t}}}

\def\rmv{{\mathrm{v}}}
\def\rmw{{\mathrm{w}}}

\def\rm0{{\mathrm{0}}}

\def\kron{\otimes}

\usepackage{enumitem}
\usepackage[usenames, dvipsnames]{color}

\newcommand{\caN}{\mathcal{N}}
\newcommand{\comm}{\mathrm{c}}

\newcommand{\crb}{\mathrm{CRB}}

\newcommand{\cvx}[1]{\mathrm{conv}\left( #1 \right)}

\newcommand{\dmse}{\mathrm{DMMSE}}
\newcommand{\Es} {{\mathcal{E}_{\mathrm{s}}} }

\newcommand{\e}[1]{{\mathbb E}\left[ #1 \right]}
\newcommand{\me}{\mathrm{e}}
\newcommand{\eff}{\mathrm{eff}}

\newcommand{\jm}{\mathrm{j}}

\newcommand{\mrad}{\mathrm{r}}

\newcommand{\mcom}{\comm}
\newcommand{\mr}{\mrad}

\newcommand{\mmse}{\mathrm{MMSE}}

\newcommand{\Ndwel}{N_\mathrm{CPI}}

\newcommand{\nm}{\mathrm{n}}

\newcommand{\mrx}{\mathrm{RX}}

\newcommand{\scnr}{\mathrm{\zeta}_{\mr}}
\newcommand{\snap}{\eta}

\newcommand{\TD}{T_{\mathrm{D}}}
\newcommand{\Ts}{T_{\mathrm{s}}}
\newcommand{\Tc}{T_{\mathrm{s}}}

\newcommand{\Tr}[1]{\mathrm{Tr}\left[ #1 \right]}

\newcommand{\txm}{\mathrm{TX}}

\newcommand{\thm}{\mathrm{th}}

\newcommand{\wa}{\mathrm{W}}
\newcommand{\na}{\mathrm{N}}

\usepackage{graphicx}

\title{Adaptive Virtual Waveform Design for Millimeter-Wave Joint Communication-Radar}
\author{{Preeti~Kumari, {\it Student Member, IEEE} Sergiy~A.~Vorobyov, {\it Fellow, IEEE}, and~Robert~W.~Heath,~Jr., {\it Fellow, IEEE}}
\thanks{ Preeti Kumari and Robert W. Heath Jr. are with the Wireless Networking and Communications Group, the University of Texas at Austin, TX 78712-1687, USA (e-mail: \{preeti\_kumari, rheath\}@utexas.edu). Sergiy~A.~Vorobyov is with the Aalto University, Konemiehentie 2, 02150 Espoo, Finland (email: sergiy.vorobyov@aalto.fi).

This research was partially supported by the U.S. Department of Transportation through the Data-Supported Transportation Operations and Planning (D-STOP) Tier 1 University Transportation Center and by the Texas Department of Transportation under Project 0-6877 entitled Communications and Radar-Supported Transportation Operations and Planning (CAR-STOP). This paper was presented in part at the IEEE ICASSP Conference, March 2017~\cite{KumNguHea:Performance-trade-off-in-an-adaptive:17} and at the IEEE ICASSP conference, April 2018~\cite{KumVorHea:VIRTUAL-PULSE-DESIGN:18}.}}

{}

\begin{document}

\maketitle
\begin{abstract}
Joint communication and radar (JCR) waveforms with fully digital baseband generation and processing can now be realized at the millimeter-wave (mmWave) band. Prior work has proposed a mmWave wireless local area network (WLAN)-based JCR that exploits the WLAN preamble for radars. The performance of target velocity estimation, however, was limited. In this paper, we propose a virtual waveform design for an adaptive mmWave JCR. The proposed system transmits a few non-uniformly placed preambles to construct several receive virtual preambles for enhancing velocity estimation accuracy, at the cost of only a small reduction in the communication data rate. We evaluate JCR performance trade-offs using the Cramer-Rao Bound (CRB) metric for radar estimation and a novel distortion minimum mean square error (MMSE) metric for data communication. Additionally, we develop three different MMSE-based optimization problems for the adaptive JCR waveform design. Simulations show that an optimal virtual (non-uniform) waveform achieves a significant performance improvement as compared to a uniform waveform. For a radar CRB constrained optimization, the optimal radar range of operation and the optimal communication distortion MMSE (DMMSE) are improved. For a communication DMMSE constrained optimization with a high DMMSE constraint, the optimal radar CRB is enhanced. For a weighted MMSE average optimization, the advantage of the virtual waveform over the uniform waveform is increased with decreased communication weighting. Comparison of MMSE-based optimization with traditional virtual preamble count-based optimization indicated that the conventional solution converges to the MMSE-based one only for a small number of targets and a high signal-to-noise ratio.

\end{abstract}

\IEEEpeerreviewmaketitle

\section{Introduction}



Millimeter-wave (mmWave) spectrum is an enabling technology to realize high data rate communication and high resolution radar sensing for many demanding applications, such as autonomous driving~\cite{ChoVaGon:Millimeter-Wave-Vehicular-Communication:16}. Traditional mmWave radars employ heavy analog pre-processing due to the use of low-speed analog-to-digital converters (ADC) and frequency-modulated continuous-wave (FMCW) technology \cite{PatTorWan:Automotive-Radars:-A-review:17}.  While effective for initial implementations, analog designs are restrictive and limit the performance as well as flexibility for futuristic radar designs~\cite{Kun:The-EU-project-MOSARIM:-A-general:12}. To address these concerns, mmWave communications hardware with high-speed ADCs can be leveraged to realize a mmWave radar system with fully digital time-domain baseband processing~\cite{BouAhmGue:PMCW-waveform-and-MIMO:16}. Further improvements can be achieved by combining radar and communication into a single joint mmWave system that uses a common waveform to enable hardware reuse.  These new joint mmWave waveforms will provide advantages in terms of cost, size, power consumption, spectrum usage, and adoption of communication-capable vehicles.

The prior approaches on joint mmWave systems are mainly classified as either joint radar-communication waveforms or joint communication-radar (JCR) waveforms. With joint radar-communication waveforms, the communication messages are modulated on top of the radar waveforms, such as pulse position modulation in \cite{BocLoyLet:A-multifunctional-60-GHz-system:10}, or phase modulation in \cite{DokShaSti:Multicarrier-Phase-Modulated:18}. These waveforms, however, do not support high data rates as the communication signal must be spread to avoid disturbing the radar properties. 

Recently, a number of mmWave JCR waveforms have been proposed that leverages consumer wireless technologies~\cite{KumGonHea:Investigating-the-IEEE-802.11ad:15,GroLopVen:Opportunistic-Radar-in-IEEE:18,KumChoGon:IEEE-802.11ad-based-Radar::17,MunMisGue:Beam-Alignment-and-Tracking:17,DokMysMis:A-mmWave-Automotive-Joint:19}. In \cite{KumGonHea:Investigating-the-IEEE-802.11ad:15}, a full-duplex IEEE 802.11ad-based radar was proposed that exploits the preamble of a single-carrier (SC) physical (PHY) layer frame to simultaneously achieve cm-level range resolution and Gbps data rates. A major limitation in \cite{KumGonHea:Investigating-the-IEEE-802.11ad:15} is that the performance of the velocity estimator was poor due to the short length of the preamble. In \cite{GroLopVen:Opportunistic-Radar-in-IEEE:18}, an opportunistic radar was developed using an IEEE 802.11ad control PHY packet, which contains a longer preamble than SC PHY, for a single target scenario.  Unfortunately, the probing signal duration is still small leading to poor velocity estimation, the data rate supported is at most 27.5 Mbps, and the update rate is very low. To enhance velocity estimation resolution, \cite{KumChoGon:IEEE-802.11ad-based-Radar::17} investigated the possibility of increasing radar integration time by developing velocity estimation algorithms that exploit multiple fixed-length IEEE 802.11ad SC PHY frames. Similar velocity enhancement techniques were used in \cite{MunMisGue:Beam-Alignment-and-Tracking:17} that proposed an IEEE 802.11ad media access control configuration to accommodate radar operations for vehicle-to-infrastructure applications. In~\cite{DokMysMis:A-mmWave-Automotive-Joint:19}, an OFDM mmWave waveform was proposed for a bi-static automotive JCR system that also exploited preambles from multiple frames at a constant spacing for enhancing velocity estimation performance. The approaches in \cite{KumChoGon:IEEE-802.11ad-based-Radar::17,MunMisGue:Beam-Alignment-and-Tracking:17,DokMysMis:A-mmWave-Automotive-Joint:19}, however, require increasing the total preamble duration to achieve desirable high-accuracy velocity estimation, which would incur degradation in the communication data rate. 

In this paper, we use sparse sensing techniques in the time domain to optimize the trade-off between communication and radar performance for the waveform design of a JCR system. We vary the frame lengths such that their preambles, which we exploit as radar pulses, are placed in a non-uniform fashion and their locations form a restricted difference basis \cite{LinSudTol:Difference-bases-and-sparse:93}. Then, we use a few non-uniformly placed pulses in a coherent processing interval (CPI) to construct a virtual difference co-waveform with several uniform virtual preambles. This virtual increase in the radar pulse integration time enables enhanced velocity estimation and a relaxed trade-off with the communication rate as compared to a uniform waveform~\cite{KumChoGon:IEEE-802.11ad-based-Radar::17}. The virtual pulse approach is conceptually similar to the staggered pulse repetition intervals (PRI) used in the classical long-range radar \cite[Ch.~17]{skolnik1990radar} and sparse sampling/arrays used in the undersampled frequency/angle/channel estimation \cite{LinSudTol:Difference-bases-and-sparse:93,VaiPal:Sparse-Sensing-With:11, MisEld:Sub-Nyquist-channel-estimation:17}. Most of the existing sparse sensing approaches, however, are focused on optimizing the sparse antenna array configurations by maximizing the antenna aperture for a given number of antenna elements. In this paper, however, we design a virtual JCR waveform using a novel minimum mean square error (MMSE)-based optimization that accurately quantifies the trade-off between communication and radar performance.
 
We make the following key assumptions in our proposed mmWave JCR waveform design. First, we assume that the location and relative velocity of a target remain constant during a coherent processing interval (CPI). This is justified by the small enough acceleration and velocity of a target relative to the radar sensor, as found in automotive applications \cite{rohling2001waveform}. Second, we assume full-duplex radar operation due to the recent development of systems with sufficient isolation and self-interference cancellation \cite{estep2014magnetic,LiJosTao:Feasibility-study-on-full-duplex:14}. Third, we assume perfect data interference cancellation on the training part of the received JCR waveform because the transmitted data is known at the radar receiver. These assumptions are explained in more detail in Section~\ref{sec:System}.

The main contributions of this paper are summarized as follows.
\begin{itemize}
\item We propose a novel formulation for a wideband JCR system that transmits virtual waveform at the mmWave band. This formulation captures the nuances of the frequency-selective sparse mmWave channel description for both communication and radar systems.  Additionally, we develop a generic virtual JCR waveform structure in the system model that can be further tuned to achieve optimal JCR performance using sparse sensing techniques.

\item We develop a novel effective distortion minimum mean square error (DMMSE) metric for communication that is comparable with the radar Cramer-Rao bound (CRB) metric for velocity estimation. The MMSE-based metrics enable us to accurately quantify the trade-off between communication and radar systems.

\item We formulate three different optimization problems for designing an adaptive JCR waveform that meets the Pareto-optimal bound. The first one minimizes the radar CRB under the constraint of a minimum communication DMMSE. The second one minimizes the communication DMMSE for a given minimum radar CRB. The third one considers a more flexible weighted average of communication and radar performance for the JCR system.
  
\item We solve the proposed JCR optimization problems for a uniform waveform and for virtual waveforms that can be represented in closed-form and contain no holes in their corresponding difference co-waveforms, such as nested virtual waveforms or Wichmann virtual waveforms. The use of specific virtual waveform configurations reduces the computational complexity for finding the optimal JCR waveform design. 

\item We carry out simulations to study the performance trade-offs in the JCR waveform design and compare the optimal performances achieved by different JCR waveform solutions. We explore the effects of signal-to-noise ratio (SNR), the number of preambles used, and the number of radar targets on the virtual waveform design. The simulation parameters are based on automotive applications and the IEEE 802.11ad-based standard. The results suggest that virtual waveforms are highly desirable at high SNR with low target density (i.e., a small ratio of target count to the number of preambles) and at low SNR with high communication DMMSE. Comparison of MMSE-based optimization with more traditional virtual preamble (VP) count-based optimization indicates that the traditional solution converges to the MMSE-based one only at low target density and high SNR. 
\end{itemize}

The work in this paper is a significant extension of our previous conference contributions \cite{KumNguHea:Performance-trade-off-in-an-adaptive:17, KumVorHea:VIRTUAL-PULSE-DESIGN:18}. In addition to a detailed exposition on the adaptive JCR waveform design, we have included a multi-target radar model and a frequency-selective communication channel model for demonstrating the superiority of virtually placed preambles as compared to uniformly placed preambles.

The rest of the paper is organized as follows. We formulate an integrated model for our proposed JCR system in Section~II. For the proposed system, we develop a radar processing technique in Section~III. Then, we describe the performance metrics and the associated trade-off for the JCR waveform design in Section~IV. In Section~V, we develop three optimization problems for the adaptive JCR waveform design. In Section~VI, we outline the main idea of virtual waveform design and different specific solution approaches for waveform optimization. We describe the simulation results and performance evaluations in Section~VII. Finally, we conclude our work and provide directions for future work in Section~VIII. 

\textbf{Notation:} We use the following notation throughout the paper: The notation
$\mathcal{N_C}(u,\sigma^2)$ is used for a complex Gaussian random variable with mean $u$ and variance $\sigma^2$. The projection matrix onto the null space of matrix $\bA$ is defined as $\mathbf{\Pi}^{\perp}_\bA = \bI - \bA (\bA^*\bA)^{-1}\bA^*$. The operators $\vert \cdot \vert$ represents the cardinality of a matrix, $\cvx{\cdot}$ denotes the convex hull, and $\Tr{\cdot}$ indicates the trace of a square matrix. The notation $(\cdot)^\rmT$, $(\cdot)^*$, and $(\cdot)^\rmc$ stand for transpose, Hermitian transpose, and conjugate of a matrix/vector, while $(\cdot)^{-1}$ represent the inverse of a square full-rank matrix. Additionally, $\mathrm{vec}(\cdot)$ vectorizes a matrix to a long vector, $\mathrm{diag}(\cdot)$ forms a vector into a diagonal matrix, while $\odot$ and $\kron$ represent Khatri-Rao and Kronecker product of matrices.

\section{System model}
\label{sec:System}
\begin{figure}[!t]
\centering
\includegraphics[width=0.7\columnwidth]{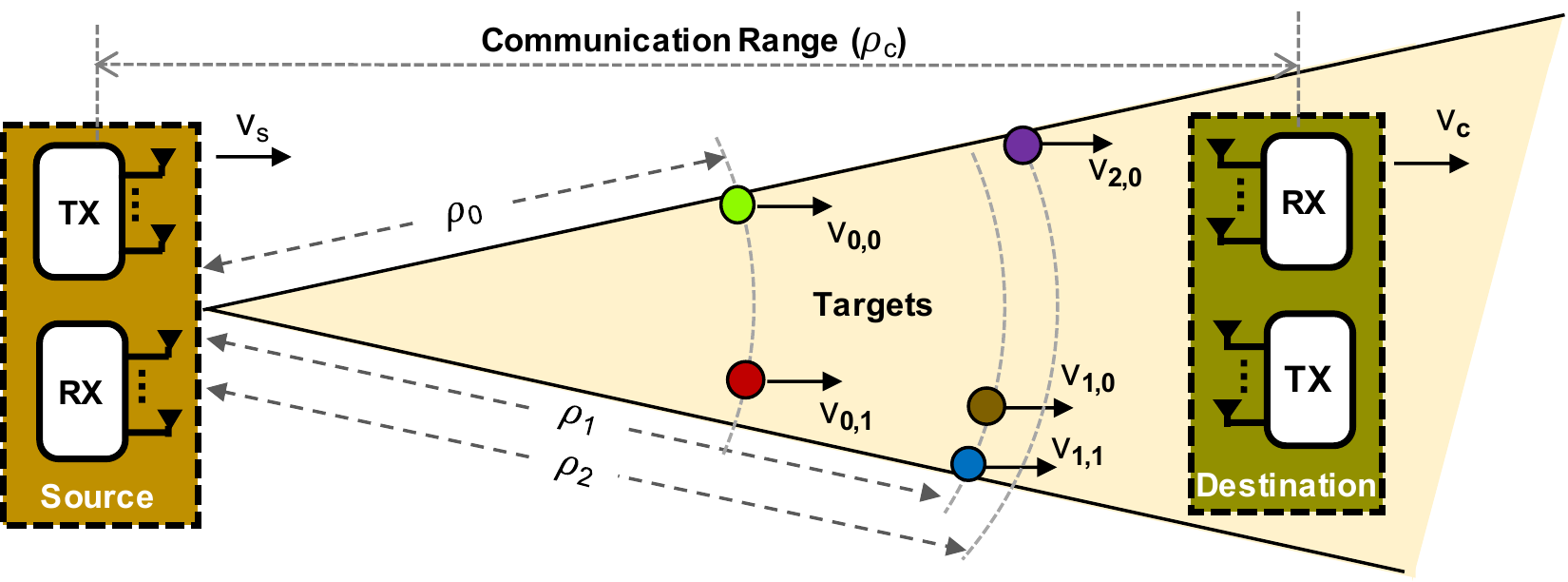}
 \caption{The source sends a mmWave waveform to the destination receiver and uses the echoes from multiple moving targets (including the destination) to estimate their ranges $\{\rho_\ell\}_{\ell=0}^{L-1}$ and velocities  $\{v_{\ell,k}\}_{\ell=0, k = 0}^{L-1, K-1}$.}
\label{fig_sysModel}
\vspace{-1.em}
\end{figure}

In this section, we formulate transmit (TX) and receive (RX) signal models for the proposed adaptive mmWave JCR system, as illustrated in Fig.~\ref{fig_sysModel}. We consider the case where a full-duplex source transmits the JCR waveform at a carrier wavelength $\lambda$ to a destination receiver at a distance $\rho_\comm$ moving with a relative velocity $v_\comm$, while simultaneously receiving echoes from the surrounding moving targets. First, we propose an adaptive single-carrier mmWave waveform structure that serves as the TX signal at the source for both communication and radar systems simultaneously. Then, we develop RX signal models for the communication receiver at the destination and the radar receiver at the source for the frequency-selective channels.

\subsection{Transmit signal model}
 \begin{figure}[!h]
\centering
\includegraphics[width=0.7\columnwidth]{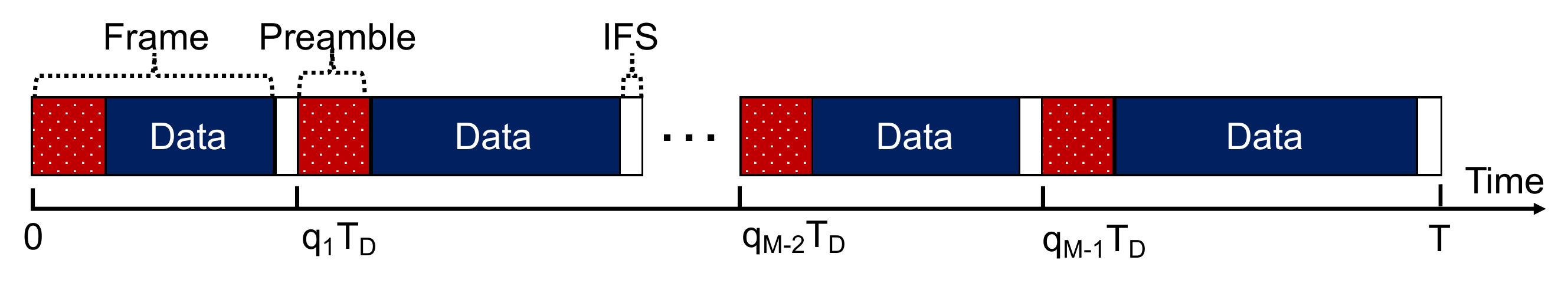}
 \caption{A CPI of $T$ seconds duration with $M$ JCR frames. Each frame contains a fixed-length preamble, a varying length data segment, and an inter-frame space (IFS) of constant duration. The length of each frame is an integer multiple, $q_m$, of the Nyquist sampling interval in Doppler domain, $T_\rmD$.}
\label{fig:GeneralFrame}
\vspace{-1.em}
\end{figure}

We consider a generic TX waveform structure with $\mu$ fraction of communication symbols and $(1-\mu)$ fraction of preamble symbols in a coherent processing interval (CPI) of $T$ seconds with $M$ frames. Each frame consists of a fixed preamble duration with a variable data length, which leads to a varying frame length as shown in Fig.~\ref{fig:GeneralFrame}. The IEEE 802.11ad standard can realize this multi-frame approach using the block/no acknowledgment policy during the communication between a dedicated pair of nodes in the data transmission interval~\cite[Ch. 9]{ieee2012wireless}. To unambiguously estimate a maximum relative target velocity $v_\mathrm{max}$ in a CPI, the $m^\thm$ frame is considered to be located at an integer multiple, $q_m$, of the Doppler Nyquist sampling interval, $T_\rmD \leq \lambda /(4 v_\mathrm{max})$. 


We denote the unit energy TX pulse-shaping filter as $g_\txm(t)$, the signaling bandwidth as $W$, and the symbol period as $\Tc \approx 1/W$.  The transmitted symbol sequence corresponding to the $m^\thm$ frame with $N_m$ symbols is denoted by $s_m[n]$, which satisfies the average power constraint $\e{\vert s_m[n] \vert^2 } = \Es$. Then, the generic complex-baseband continuous-time representation of the single-carrier TX waveform in a CPI is given as
\begin{equation} \label{eq:TXCont}
x(t) = \sum_{m =0}^{M -1}  \sum_{n=0}^{N_m-1} s_m[n] g_\txm(t -n\Tc-q_mT_\rmD) .
\end{equation}
The generic TX waveform parameters, such as the location and size of the $m^\thm$ frame, can be further optimized to achieve desirable JCR performance as described in Section~\ref{sec:Problem}.


In this paper, we consider a single data stream model that supports analog beamforming with frequency flat TX/RX beam steering vectors~\cite{KumChoGon:IEEE-802.11ad-based-Radar::17}. We assume that the source and destination align their beams toward each other with line-of-sight (LoS) frequency-selective communication and radar channel. The insights and analysis of this work can also be extended to mmWave multiple-input-multiple-output (MIMO) radar research using low-resolution ADCs \cite{KumMazMez:Low-Resolution-Sampling-for-Joint:18} by focusing on the signal model for a given angular bin. We will now formulate the JCR received signal model after the TX/RX beamforming for multiple frames in a CPI.

\subsection{Receive signal models}

We consider a dwell time consisting of $\Ndwel$ coherent processing intervals. During the dwell time, we assume that the acceleration and the relative velocity of a moving target is small enough to assume constant velocity and that the target is quasi-stationarity (constant location parameters). We assume a frequency-selective Rayleigh fading model for both communication and radar channels during the dwell time for simplicity. This work can also be extended for a general Ricean fading mmWave channel model with block sparsity by deriving corresponding CRB bounds for different Ricean fading factors that lies between zero (Rayleigh fading) and infinity (no fading). In each of the CPIs with $M$ frames, we consider a block fading model that assumes a constant channel gain for each delay tap. 
\subsubsection{Communication received signal model}
For evaluating the trade-off between communication and radar performance, we model the communication signal received at a distance $\rho_\comm$. Assuming a highly directional mmWave LoS communication link is established between the source and destination, the large-scale path-loss $G_\comm$ is given as~\cite{RapMacSam:Wideband-Millimeter-Wave-Propagation:15} 
\begin{equation}
G_\comm = \frac{G_\txm G_\mrx  \lambda^2 } {(4 \pi)^2 \rho_\mcom^\mathrm{PL}},
\end{equation} 
where $\mathrm{PL}$ is the path-loss exponent, $G_\txm$ is the TX array gain, and $G_\mrx$ is the RX array gain. 

After TX/RX beamforming, symbol synchronization, and frequency synchronization, the received communication signal is a sum of the contributions from $L_\mcom+1$ delayed and attenuated copies of the transmitted signals as well as the additive white Gaussian noise (AWGN) with zero mean and variance $\sigma_{\nm}^2$. The small-scale complex gain of the $\ell^\thm$ channel delay path, $\alpha_\mcom[\ell]$, is assumed to be independently and identically distributed (IID) $\mathcal{N_C}(0,\sigma^2_\mcom[\ell])$, where $\sigma^2_\mcom[\ell]$ represents the average tap power normalized such that $\sum_{\ell=0}^{L_\mcom}  \sigma^2_\mcom[\ell]  = 1$. Therefore, the communication RX signal $y_{\mcom,m}[n]$ corresponding to the $n^\thm$ symbol in the $m^\thm$ frame with noise $w_{\comm,m}[n]$ is
\begin{equation} \label{eq:commRx}
y_{\mcom,m}[n] = \sqrt{G_\mcom} \sum_{\ell = 0}^{L_\mcom} \alpha_\mcom[\ell]s_m[n-\ell]+w_{\mcom,m}[n] .
\end{equation}
The communication SNR is defined as $\zeta_\mcom \triangleq \Es G_\mcom/\sigma_\nm^2$.

\subsubsection{Radar received signal model}
We represent the doubly selective (time- and frequency-selective) mmWave radar channel obtained after TX/RX beamforming using the virtual representation obtained by uniformly sampling in the range dimension \cite{KumChoGon:IEEE-802.11ad-based-Radar::17}.  We assume that there are $L$ range bins. The $\ell^\thm$ range bin is assumed to consist of a few, $K[\ell]$, virtual scattering centers with different velocities. Each of the $(\ell,k)^{\mathrm{th}}$ virtual scattering center is characterized by its distance $\rho[\ell]$, delay $\tau[\ell]$, velocity $v_{k}[\ell]$, Doppler shift $\nu_{k}[\ell] = 2 v_{k}[\ell]/\lambda$, and radar cross-section $\sigma_{\mathrm{RCS},k}[\ell]$.
The channel gain, $\beta_{k}[\ell]$, corresponding to the $(\ell,k)^{\mathrm{th}}$ virtual target scattering center is (as used extensively in previous work, e.g., \cite{bazzi2012estimation}) 
\begin{equation}
\beta_{k}[\ell] = \frac {G_\txm G_\mrx \lambda^2 \sigma_{\mathrm{RCS},k}[\ell]}{64 \pi^3 \rho^4[\ell]} 
\end{equation}
and is assumed IID $\mathcal{N_C}(0,G_{k}[\ell])$.

After matched filtering (MF) with the RX pulse shaping filter $g_\mrx(t)$ and symbol rate sampling, the received radar signal is a sum of the contributions from the attenuated, delayed, Doppler-shifted, and sampled MF echoes as well as the AWGN with zero mean and variance $\sigma_{\nm}^2$. Therefore, the radar received signal model corresponding to the $m^\thm$ frame with the net TX-RX pulse shaping filter $g(t) = g_\txm(t) \ast g_\mrx(t)$, delayed and sampled MF echo from the $\ell^\thm$ range bin $ \varepsilon_m[n,\ell]  = \sqrt{\Es}\sum_{i=0}^{N_m-1}s_m[i]g((n-i)\Ts-q_m \TD-\tau[\ell])$ and noise $w_{\mr,m}[n]$ is given as
\begin{equation} \label{eq:radarRX}
\begin{aligned}
y_m[n] &=   \sum_{\ell=0}^{L-1}\! \varepsilon_m[n,\ell] \sum_{k=0}^{K[\ell]-1} \! \sqrt{ \beta_{k}[\ell]} \me^{- \jm 2\pi \nu_{\ell,k} (n\Ts + q_mT_\rmD)} & + w_{\mr,m}[n].
\end{aligned}
\end{equation}
The received echo $y_m[n]$ is comprised of reflections corresponding to the TX symbol $s_m[n]$ as well as an intersymbol interference from the other TX symbols that depends on the choice of the TX/RX pulse shaping filters and the doubly-selective radar channel parameters. 


We use the training sequences for radar parameter estimation due to to their good correlation properties. The training part might incur some interference from the data part because of the larger delay spread observed in the two-way radar channel as compared to the guard interval employed between the preamble and the data part in a typical communication system. Motivated by the recent development of non-orthogonal multiple access techniques with successive interference cancellation \cite{DaiWanYua:Non-orthogonal-multiple-access:15}, we assume perfect cancellation of the data part on the received training signal. Developing and evaluating algorithms to cancel the communication data interfence while receiving the radar segment reflection is a subject of future work. 

We assume the channel to be time invariant within the preamble duration of a single frame due to slow enough velocity and small enough preamble duration. Therefore, the received signal model corresponding to the training part with $\varepsilon_{\rmt}[n,\ell]$ as the preamble/training part of $\varepsilon[n,\ell]$ that remains same for each frame is 
\begin{equation} \label{eq:FinalRadarRX}
\begin{aligned}
y^{\rmt}_m[n] &=  \sum_{\ell=0}^{L-1} \varepsilon_{\rmt}[n,\ell] \sum_{k=0}^{K[\ell]-1} \!  \sqrt{\beta_{k}[\ell]} \me^{- \jm 2\pi \nu_{k}[\ell] q_mT_\rmD} 
+ w_{\mrad,m}[n].
\end{aligned}
\end{equation}
Note that the virtual channel model with $\sum_{\ell=0}^{L-1} K[\ell]$ scattering centers is used in \eqref{eq:FinalRadarRX}.


\section{Proposed radar processing}
In this section, we propose a radar processing technique for estimating target velocities using the proposed JCR frames of the same or varying lengths. The radar processing technique exploits the preamble part of the JCR frame that consists of training sequences with good auto-correlation properties, such as Golay complementary sequences used in the IEEE 802.11ad-based automotive radar applications~\cite{KumChoGon:IEEE-802.11ad-based-Radar::17,GroLopVen:Opportunistic-Radar-in-IEEE:18}. First, we estimate the channel using a typical communication-based preamble processing algorithm that exploits the correlation properties of the training sequences~\cite{KumChoGon:IEEE-802.11ad-based-Radar::17}. Then, we calculate the target velocities from the estimated channel using super-resolution velocity estimation algorithms. 

Denoting $b_{k}[\ell]\triangleq \gamma \sqrt{\Es \beta_{k}[\ell]}$ as signal amplitude in the channel, $\gamma$ as the integration gain due to the correlation-based channel estimation, ${w}_m[\ell]$ as the noise in the channel, and $u_{k}[\ell] \triangleq \nu_{k}[\ell] \TD$ as the discrete Doppler frequency,  the channel corresponding to the detected target in the $\ell^\thm$ range bin that is derived using the $m^\thm$ frame received in (\ref{eq:FinalRadarRX}) is given as
\begin{equation} \label{eq:radChannel}
{h}_m[\ell] = \sum_{k=0}^{K[\ell]-1}  b_{k}[\ell] \me^{- \jm 2 \pi u_{k}[\ell]q_m} +  w_m[\ell],
\end{equation} 
where ${w}_m[\ell]$ is distributed as $\mathcal{N_C}(0,\sigma_{n}^2)$. The channel vector corresponding to the $\ell^\thm$ range bin for a CPI of $M$ frames is ${\mathbf{h}}[\ell] \triangleq \left[ {h}_0[\ell], {h}_1[\ell], \cdots, {h}_{M-1}[\ell] \right]^\rmT$. 

The focus of this paper is on target velocity estimation. Therefore, we describe algorithms for estimating the velocity of a single target in a specific range bin, say $\ell_0$, which can be similarly performed for other range bins. To simplify the expressions, we henceforth omit the notation $\ell$ denoting the range bin (e.g., $b_{k}[\ell]$ becomes just $b_k$). 

Denoting the channel signal amplitude vector ${\mathbf{b}} \triangleq [ {b_{0}}, {b_{1}},\cdots, b_{K-1} ]^\rmT$, the channel Doppler vector corresponding to the $k^\thm$ velocity, $\mathbf{d}(v_k) \triangleq [ 1, e^{j2 \pi u_k q_1}, \cdots, e^{j 2 \pi u_k q_{M-1}}]^\rmT$, the channel Doppler matrix ${\mathbf{D}} \triangleq [ \mathbf{d}(v_{0}), \mathbf{d}(v_{1}), \cdots, \mathbf{d}(v_{K-1} )]$, and the channel noise vector ${\mathbf{w}} \triangleq [ {w_{0}}, {w_{1}},\cdots, \\ w_{K-1} ]^\rmT$, the channel vector corresponding to the range bin $\ell_0$ with $K >0$ detected targets is given by
\begin{equation} \label{eq:h_mul}
\mathbf{h} = {\mathbf{D}} \mathbf{b}+ \mathbf{w}.
\end{equation}
This channel vector is used for target velocity estimation.

Due to space limitations and for simplicity of our basic study here, we focus on subspace methods, in particular on the class of multiple signal classification (MUSIC) techniques, for velocity estimation algorithms among many possible approaches \cite{stoica2005spectral,yang2016sparse,WanNeh:Coarrays-MUSIC-and-the-Cram:17}. The velocity resolution obtained by the subspace methods is not constrained by the duration of the CPI as in the fast Fourier transform (FFT)-based technique used in \cite{KumChoGon:IEEE-802.11ad-based-Radar::17}. Therefore, a subspace method can provide higher resolution in the mobile environment with limited CPI.  Subspace-based velocity estimation using multiple preambles in a CPI can be performed by exploiting sample covariance matrix of the channel in \eqref{eq:h_mul}~\cite{stoica2005spectral,yang2016sparse,WanNeh:Coarrays-MUSIC-and-the-Cram:17}.
The covariance matrix of the channel with $p_k = \e{b_k b_k^*}$ as the power of the $k^\thm$ target and $\mathbf{P} \triangleq \mathrm{diag}(p_0,p_1,\cdots,p_{K-1})$ as the target covariance/power matrix is given by
\begin{equation} \label{eq:radCovar}
\bR = \bD \mathbf{P} \bD^* + \sigma^2_\nm \mathbf{I}.
\end{equation}
We define the SNR of the received radar signal at the source vehicle corresponding to the $k^\thm$ target as $\scnr[k] =  p_k /\sigma_{\nm}^2$.

We evaluate the CRB performance metric for the velocity estimation performance of the subspace method using the channel covariance matrix in \eqref{eq:radCovar}, as described in Section~\ref{sec:Perf}.  In Section~\ref{sec:Solution}, we further illustrate the MUSIC-based velocity estimation algorithms for specific waveform design solutions that exploits the finite snapshot version of \eqref{eq:radCovar}.

\section{Performance metrics}
\label{sec:Perf}
In this section, we first describe the spectral efficiency performance metric for communication systems and the CRB performance metric for radar systems. Then, we describe a novel metric for assessing the system trade-off between radar and communication for a joint waveform design.

\subsection{Communication performance metric}
The channel capacity for the received communication signal in (\ref{eq:commRx}) with eigenvalues of the channel matrix as $\{ \lambda_\mcom[n] \}_{n=1}^N$, fraction of data symbols $\mu = 1$, data power coefficients $\{ \xi[n] \}_{n=1}^{N}$ satisfying $\frac{1}{N} \sum_{n = 1}^{N} \xi[n]$ = 1, data symbol $s[n] \sim \caN(0,\xi[n]\Es)$, and noise $w_\mcom[n] \sim \caN(0,\sigma^2_\rmw)$ is obtained by allocating optimum power based on the vector coding among a block of $N \to \infty$ symbols of the single-carrier waveform \cite[Ch.~4]{HeaLoz:Foundations-of-MIMO-Communication:19}. The maximum communication spectral efficiency achieved using vector coding transmission for $N \to \infty$ is expressed as
\begin{equation} \label{eq:spec}
r =  \frac{1}{N}\sum_{n = 1}^{N} \log_2 \left( 1 + {\zeta_\mcom \lambda_\mcom[n] \xi[n]}\right) \text{\, bits/s/Hz}
\end{equation}
and the channel capacity in bits per second (bps) is given as $C = Wr$.
Note that the achievable spectral efficiency of a communication system depends on the implemented precoder and equalizer~\cite{TakKyrHan:Performance-evaluation-of-60-GHz-radio:12,liu2013digital}, and are all upper bounded by (\ref{eq:spec}) for $N \to \infty$.


When $\mu \leq 1$ fraction of communication symbols are transmitted in a CPI with the channel capacity $C$, we define the effective communication data rate as $C_\eff = \mu C$, as in~\cite[Ch.~7]{HeaLoz:Foundations-of-MIMO-Communication:19}. Additionally, we can define the effective communication spectral efficiency for $\mu \leq 1$ as 
\begin{equation} \label{eq:specEff}
r_\eff = \mu r =  { \frac{\mu}{N}\sum_{n = 1}^{N} \log_2 \left( 1 + {\zeta_\mcom \lambda_\mcom[n] \xi[n]}\right)} \text{\, bits/s/Hz},
\end{equation}
which satisfies $C_\eff = W r_\eff$ bps when $N \to \infty$.
 

\subsection{Radar performance metric}
The CRB is a lower bound on the variance of an unbiased estimator. For white Gaussian noise, the CRB is also a lower bound on the MMSE for radar parameter estimation and is used for asymptotic analysis. To express the CRB corresponding to (\ref{eq:radCovar}) with $\eta$ snapshots, we denote the derivative of the channel Doppler matrix $\mathbf{D}$ with respect to the $K$ velocity parameters $\{ v_k \}_{k=1}^K$ as $\dot{ \mathbf{D}}$, the co-waveform Doppler matrix $\bD_\rmq \triangleq \bD^\rmc \odot \bD$, the derivative of the co-waveform Doppler matrix as $\dot{\bD}_\rmq \triangleq\dot{ \mathbf{D}}^* \odot \mathbf{D} + \mathbf{D}^* \odot \dot{ \mathbf{D}}$, identity vector $\bi \triangleq \mathrm{vec}\{\bI\}$. Then, the CRB matrix for the $K$ estimated velocities with $\bE \triangleq  (\bR^\rmT \kron \bR)^{-1/2} \dot{\bD}_\rmq \mathbf{P} $ and $\bF \triangleq (\bR^\rmT \kron \bR)^{-1/2} [\bD_\rmq \quad \bi]$ is~\cite{WanNeh:Coarrays-MUSIC-and-the-Cram:17}
\begin{equation} \label{eq:radCRB}
\crb=\frac{1}{\snap} \left( \bE^* \mathbf{\Pi}^{\perp}_{\bF} \bE \right)^{-1}.
\end{equation}
Note that the use of multiple snapshots can be achieved by using multiple CPIs within a dwell time. The number of snapshots depends on the dwell time and the latency desired for radar parameter estimation. When the Fisher information matrix is necessarily singular, the CRB does not exist implying that no unbiased estimator with finite variance exists \cite{StoMar:Parameter-estimation-problems:01}.



\subsection{Joint communication-radar performance metric}
In this section, we develop a novel JCR metric to quantify the trade-off between the radar and the communication performance.
In \cite{Bli:Cooperative-radar-and-communications:14}, a range estimation rate metric for radar was proposed and is analogous to the data rate used in communication systems. The radar estimation, however, is not drawn from a countable distribution similar to communication data symbols. Therefore, the estimation rate metric is not an accurate representation of radar performance. The derivation of estimation rate for radar round-trip delay is also not easily extendable to other radar parameters because several underlying simplifications made in \cite{Bli:Cooperative-radar-and-communications:14,HerBli:Spectrum-management-and-advanced:18} may become invalid for the estimation of these other parameters \cite{PauBli:Extending-joint-radar-communications:15}. Additionally, the number of radar performance metrics (e.g., range/velocity/direction of multiple targets, number of detectable targets, probability of detection and false alarm, range/velocity/angular resolution) that depend on $\mu$ is much larger than the few performance metrics used in communication. Therefore, instead of deriving equivalent estimation or information rates for each of these radar parameters in different scenarios, as in \cite{PauBli:Extending-joint-radar-communications:15}, we propose an effective communication DMMSE metric similar to a radar CRB performance metric.

To formulate an effective scalar communication metric, which parallels the concept of the radar CRB for JCR waveform design optimization, we propose an MMSE-based communication metric analogous to the distortion metric in the rate-distortion theory \cite[Ch. 10]{CovTho:Elements-of-information-theory:12}. The MMSE of a communication system with net spectral efficiency $r$, $\mu = 1$, and $i^\thm$ spectral efficiency $r_i =  \log_2(1 + \zeta_\mcom \lambda_\mcom[n] \xi[n])$ in (\ref{eq:specEff}) is given as~\cite{PalVer:Gradient-of-mutual-information:06}
\begin{equation} \label{eq:MMSEdef}
\begin{aligned}
\mmse_\comm &= {\e{(\bs-\hat{\bs}(\by))((\bs-\hat{\bs}(\by)))^*}} \\ 
&= \mathrm{diag}{\left( 2^{-r_1} , \cdots, 2^{-r_N}\right)} .
\end{aligned}
\end{equation}
 
Using (\ref{eq:spec}) and (\ref{eq:MMSEdef}), the relationship between $\mmse_\comm$ and $r$ becomes
\begin{equation}
\frac{1}{N}\Tr {\log_2 {\mmse_\mcom}} = -r .
\end{equation}
Therefore, the effective communication DMMSE that satisfies 
\begin{equation} \label{eq:effMMSE_Det}
\frac{1}{N}\Tr {\log_2 {\dmse}} =  -r_\eff = -\mu \cdot r
\end{equation} 
is given as
\begin{equation} \label{eq:effMMSE}
\dmse =  \mmse_\comm^\mu.
\end{equation} 
The effective DMMSE derived for a single-target scenario in \cite{KumNguHea:Performance-trade-off-in-an-adaptive:17} also follows relations similar to (\ref{eq:effMMSE_Det}) and (\ref{eq:effMMSE}). Note that determinant and largest eigenvalues could be used instead of trace in (\ref{eq:effMMSE_Det}). Indeed, the determinant is a volume, the largest eigenvalue is the length along the longest access, and the trace is a sum of all the eigenvalues. Therefore, the trace is a reasonable selection for a MMSE-based JCR performance metric. The performance trade-off between communication and radar can then be quantified in terms of the following scalar quantities: $\frac{1}{N}\Tr{\log \dmse_\eff}$ and $\frac{1}{K}\Tr{\log \crb}$. Since the communication DMMSE and the radar CRB values are usually substantially different, the log-scale is used to achieve proportional fairness (PF) similar to the problem of resource allocation in multi-user communication \cite[Ch. 7]{HeaLoz:Foundations-of-MIMO-Communication:19}.

\section{Adaptive waveform design problem formulation}
\label{sec:Problem}


The JCR performance optimization problem is a multi-objective (two-objective) problem of simultaneously optimizing both the radar performance, in terms of, for example, decreasing the velocity CRB, and communication performance, in terms of minimizing the effective communication DMMSE. The scalarization approach is known to achieve a Pareto optimal point for multiple convex objectives \cite[Ch. 4]{BoyVan:Convex-optimization:04}. Therefore, the JCR performance optimization can be formulated as the weighted average of a convex hull of communication and radar MMSE-based performance metrics. Denoting $\varphi_\mcom(\dmse_\eff) \triangleq \cvx{\Tr{ \log \dmse_\eff}}$ as the communication metric and $\varphi_\mr(\crb) \triangleq \cvx{\Tr{\log \crb}}$ as the radar metric, the JCR performance optimization problem can be formulated as
\begin{eqnarray} \label{eq:weighted}
&\underset{M,\{ q_m \}_{m=1}^M}{\text{minimize}}&\;\omega_\mr {\varphi}_\mrad(\crb)    + \omega_{\comm}  {\varphi}_\comm(\dmse_\eff)   \nonumber \\
&{\text{subject to}}& \{ T, K, \rho \} =  \mathrm{constants,} \nonumber\\
&&  0 < q_1 < \cdots < q_{M-1} < T/T_\rmD,
\end{eqnarray}
where $\omega_\mr \geq 0 $ and $\omega_{\comm} \geq 0 $ are the normalizing and weighting factors assigning the priorities for radar and communication tasks. Note that the weights can be adjusted adaptively with respect to the requirements imposed by different scenarios, such as varying radar SNR.
 
Alternatively, problem \eqref{eq:weighted} can be modified as a minimization of one of the objectives with the second written as a constraint that would guarantee an acceptable performance for one of the tasks. The radar CRB constrained formulation with a minimum required CRB $\Upsilon_\mrad$ can be expressed as
\begin{eqnarray} \label{eq:RadConst} 
&\underset{M,\{ q_m \}_{m=1}^M}{\text{minimize}}&\; {\varphi}_\comm(\dmse_\eff) \nonumber \\
&{\text{subject to}}&  {\varphi}_\mrad(\crb)  \leq \Upsilon_\mrad,\nonumber\\
&{\text{}}&  \{ T, K, \rho  \} = \mathrm{constants,} \nonumber\\
&&  0 < q_1 < \cdots < q_{M-1} < T/T_\rmD.
\end{eqnarray}
The optimization in \eqref{eq:RadConst} simplifies to finding the minimum number of frames $M$ that meets the required radar CRB value, whenever a specific sparse pulse configuration, such as coprime pulses, is assumed.

The communication DMMSE-constrained formulation with minimum required DMMSE $\Upsilon_\mcom$ is given by
\begin{eqnarray} \label{eq:CommConst} 
&\underset{M,\{ q_m \}_{m=1}^M}{\text{minimize}}&\;  {\varphi}_\mrad(\crb) \nonumber \\
&{\text{subject to}}&  {\varphi}_\comm(\dmse_\eff)  \leq \Upsilon_\mcom,\nonumber\\
&{\text{}}&  \{ T, K, \rho  \}= \mathrm{constants,} \nonumber\\
&&  0 < q_1 < \cdots < q_{M-1} < T/T_\rmD.
\end{eqnarray}
The optimization in \eqref{eq:CommConst} for a constant predefined number of frames $M$ for a large enough CPI $T$ simplifies to the optimization of frame locations.




\section{Adaptive waveform design solutions}
\label{sec:Solution}
Finding optimal virtual waveform designs as the solutions of the JCR optimization problems proposed in Section~\ref{sec:Problem} is computationally difficult (generally has combinatorial complexity). To ensure polynomial complexity for solving the JCR optimization problems and for mathematical tractability, we use specific configurations of preamble locations that have good ambiguity functions. It helps to dramatically reduce the optimization complexity and problem size to only a few variables depending on the specific configurations used. In this section, we present different adaptive single-carrier waveform designs based on the JCR optimization problem formulations along with their associated algorithms. These solutions are mainly classified as uniform or non-uniform (virtual) waveform designs. In the uniform waveform design, the preambles are placed at a Doppler Nyquist rate $1/\TD$, whereas in the non-uniform waveform design, the preambles are placed at a Doppler sub-Nyquist rate $1/(q_m\TD)$ with integer $q_m > 1$.

\begin{figure}[!t]
\begin{minipage}[h]{\columnwidth}
\centering
\includegraphics[scale=0.7]{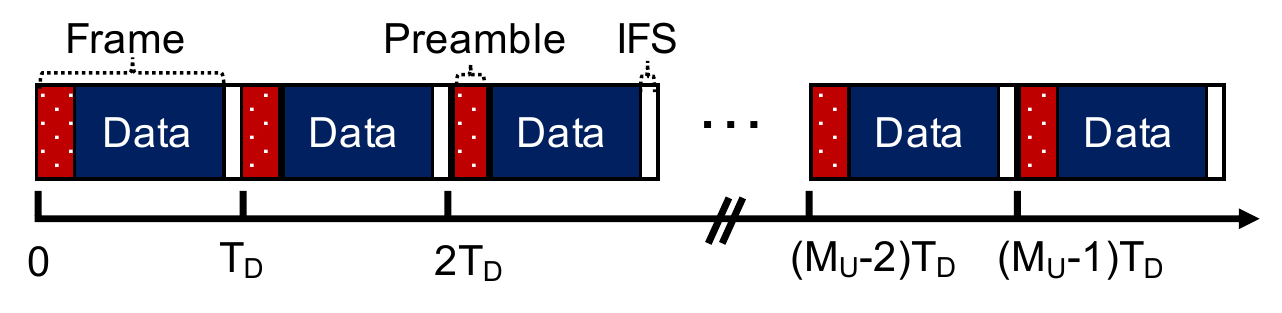}
\newline
{(a) }
   \end{minipage}
 \vfill
\begin{minipage}[h]{\columnwidth}
\centering
\includegraphics[scale=0.7]{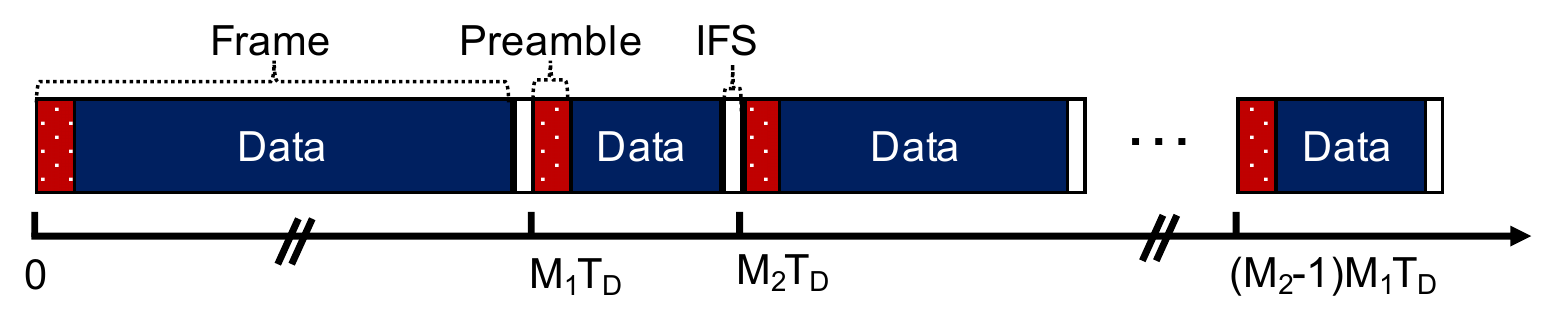}
\newline 
 {(b) } 
 \end{minipage}
   \caption{Different virtual waveforms: (a) Uniform one, where a CPI consists of $M_\rmU$ equi-spaced frames of duration $\TD$. (b) Coprime one, where a CPI consists of non-uniformly placed $M = M_1+M_2$ frames.}
  \label{fig:VirtualWaveform}
  \vspace{-.7em}
  \end{figure}

\subsection{Uniform waveform design}
In this approach, multiple frames, $M_\rmU$, are placed at a constant Doppler Nyquist sampling interval, $T_\rmD$, in a CPI of $T \geq M_\rmU T_\rmD$ seconds~\cite{KumChoGon:IEEE-802.11ad-based-Radar::17}, as shown in Fig.~\ref{fig:VirtualWaveform}(a).
Therefore, the $m^\thm$ frame in a CPI is located at $q_m = m$ and the set of all preamble locations in a CPI is $\mathcal{M}_\rmU = \{1, 2,\cdots, M_\rmU\}$. 

For uniform waveform design, the CRB expression in (\ref{eq:radCRB}) exists when the target count $K$ is smaller than the number of preambles $M_\rmU$.
Additionally, $\crb$ in (\ref{eq:radCRB}) reduces to zero as SNR goes to infinity~\cite{StoNeh:Performance-study-of-conditional:90}. We use the standard MUSIC algorithms, which is asymptotically efficient, for estimating velocity using uniform waveforms when $K < M_\rmU$ \cite{StoNeh:MUSIC-maximum-likelihood:89}.

The feasibility and behavior of optimal solutions in the waveform design problem formulations in (\ref{eq:weighted}), (\ref{eq:RadConst}), and \eqref{eq:CommConst} depends on the velocity CRB in \eqref{eq:radCRB} as well as the communication DMMSE in \eqref{eq:effMMSE}. Since the CRB in \eqref{eq:radCRB} monotonically reduces with the increasing SNR, the optimal solutions for uniform waveform designs in (\ref{eq:weighted}), (\ref{eq:RadConst}), and \eqref{eq:CommConst} will continuously improve with the decrease in the target distance $\rho$. Additionally, with decreasing $\rho$, the optimal number of preambles, $M^*_\rmU$, in (\ref{eq:RadConst}) will decrease, $M^*_\rmU$ in (\ref{eq:CommConst}) will be constant, and the change in $M^*_\rmU$ will be adapted based on the rate of decrease in the radar CRB with $\rho$ for weighted average optimization-based design in (\ref{eq:weighted}). With an increase in $K$ in JCR problem formulations, the CRB in \eqref{eq:radCRB} degrades and its existence depends on $M_\rmU$. Therefore, with growing $K$, the optimal solutions for uniform waveform designs in (\ref{eq:weighted}) as well as in (\ref{eq:RadConst}) will rapidly degrade with a steep increase in $M^*_\rmU$, while the feasibility of optimal waveform design in (\ref{eq:weighted}), (\ref{eq:RadConst}), and \eqref{eq:CommConst} will be severely limited. Additionally, the optimal solutions achieved by adaptive JCR waveform designs for all three problems are limited. This is because the use of multiple frames placed at the Nyquist rate in a CPI will lead to a large physical increase in the preamble duration, thereby significantly decreasing the communication spectral efficiency. These insights are further explained in Section~\ref{sec:Simulation}.

%




\subsection{Non-uniform waveform design}



For non-uniform waveform designs, $M_\rmV$ frames are non-uniformly placed in a CPI of $T \geq M_\rmV T_\rmD$, as shown in Fig.~\ref{fig:VirtualWaveform}(b). Therefore, the preambles are placed in a sub-Nyquist fashion with varying $N_m\Ts \geq T_\rmD$ spacing between them. Here, the $m^\thm$ frame is located at $q_m \geq m$ and the set of all preamble locations in a CPI $\mathcal{M}_\rmV$ is a sparse subset of the contiguous set $\mathcal{M} = \{1, 2, \cdots, M \}$ with $\vert \mathcal{M}_\rmV \vert = M_\rmV$ that guarantees the desirable velocity estimation performance. The location of preambles in a CPI can be chosen randomly or in a determined fashion. 

The VP locations for the non-uniform waveform is obtained using the following difference co-waveform of $\mathcal{M}_\rmV$ 
\begin{equation}
\mathcal{C}_\rmV = \{m_p - m_q \: \vert \: 1 \leq p, q \leq M_\rmV \}.
\end{equation}
Vectorizing $\bR$ yields the co-waveform signal model expressed as 
\begin{equation}
\br = \mathrm{vec}(\bR) = \bD_\rmq \boldsymbol{\bp}+\sigma^2_\rmn \bi,
\end{equation} 
where $\bD_\rmq \triangleq \bD^\rmc \odot \bD$ and $\bp \triangleq [p_1, p_2,\cdots, p_K]$. The matrix $\bD_\rmq$ also represents the steering matrix of $\mathcal{C}_\rmV$ \cite{PalVai:Nested-Arrays:-A-Novel:10}. 

For tractable analysis, we use deterministic non-uniform waveforms that can be represented in closed-form and contain no holes in their corresponding difference co-waveforms. MUSIC-like algorithms can then be applied on the full contiguous stretch of $\vert \mathcal{C}_\rmV \vert$ elements in the hole-free difference co-waveforms. The approach developed in this paper can also be extended to other sparse waveforms. 

The CRB expression in (\ref{eq:radCRB}) is valid under the condition that $2K \leq \vert \mathcal{C}_\rmV \vert$ for hole-free difference co-waveforms~\cite{KooPal:Cramer-Rao-Bounds-for-Underdetermined:16}. The cardinality of the difference co-waveform $\mathcal{C}_\rmV$ depends on the placements of non-uniform preambles $\mathcal{M}_\rmV$ and can be used to identify $\mathcal{O}(M_\rmV^2)$ sources. For $K<<M_\rmU$ and $M_\rmU = M_\rmV = M$, the CRB for non-uniform waveforms also decreases much faster than for uniform waveforms as $M$ increases \cite{WanNeh:Coarrays-MUSIC-and-the-Cram:17}. At $K>M_\rmV$, however, the CRB for the non-uniform waveform design may not reduce to zero when SNR goes to infinity. 

Since the non-uniform waveform design usually needs a lower $M_\rmV$ to achieve a given radar CRB and a much higher $M_\rmV$ to achieve a valid CRB for a given $K$ than a uniform waveform design, a non-uniform waveform design allows a larger set of feasible solutions for the JCR waveform optimization problems in (\ref{eq:weighted}), (\ref{eq:RadConst}), and \eqref{eq:CommConst}. Additionally, the lower the cardinality of the sparse set $\mathcal{M}_\rmV$, the smaller the overhead on the effective communication spectral efficiency. Therefore, non-uniform waveform design allows a reduced trade-off between the radar CRB and the communication DMMSE for low target density, thereby resulting in an enhanced optimal JCR waveform design. The saturation effect observed in non-uniform waveform design at high SNR, however, reduces the advantage of non-uniform JCR waveform design over uniform JCR waveform design at small radar distances. 

Among several redundancy waveforms with no holes \cite{RajKoi:Comparison-of-sparse-sensor:17}, Wichmann and nested waveforms are relevant. The {\it Wichmann virtual waveform (WVW)} is known to yield the largest aperture co-waveform for all redundancy waveforms with more than 8 elements \cite{LinSudTol:Difference-bases-and-sparse:93}. The inter-preamble spacings of the WVW is \cite{wichmann1963note}
\begin{equation}
d \mathcal{M}_{\wa} \!\!=\!\! \{ 1^{(p)}\!, p+1, (2p+1)^{(q)}\!, (4p+3)^{(q)}\!, (2 p +2)^{(p+1)}\!,1^{(p)}\},
\end{equation}
where $p,q \in \mathbb{N}$ and the notation $p^{(q)}$ represents $q$ repetitions of $p$.
The VP count $\vert \mathcal{C}_{\wa} \vert$ of the WVW is 
\begin{equation}
\vert \mathcal{C}_{\wa} \vert =4p(p+q+2) + 3(q+1).
\end{equation}
In most of the prior work in sparse array optimization,  $\vert \mathcal{C}_\rmv\vert$ is maximized for a given $\vert \mathcal{M}_{\rmV} \vert$. Using this traditional optimization criteria, optimum $p^*$ is the closest non-negative integer solution to $ (\vert \mathcal{M}_{\rmV} \vert -4)/{6}$ and optimum $q^* ={\vert \mathcal{M}_{\rmV} \vert -4p^*-3}$ \cite{RajKoi:Comparison-of-sparse-sensor:17}.
The fraction of communication data symbols $1 -\mu^*_{\wa}$ for the WVW in a CPI is
\begin{equation}\label{eq:alphaII}
1- \mu_{\wa}^* = 1 - \frac{(4p^*+q^*+3) P \Ts+M_{\wa}^* T_\mathrm{IFS}}{T},
\end{equation}
where $M_{\wa}^*$ is the optimal minimum $\vert \mathcal{M}_\rmV\vert$ for the WVW and $T_\mathrm{IFS}$ is the interframe spacing. In Section~\ref{sec:Simulation}, we will compare $p^*$ and $q^*$ values obtained using the VP count-based optimization and the CRB-based optimization.

The {\it Nested virtual waveform (NVW)} is widely used in MIMO radar for direction-of-arrival estimation \cite{PalVai:Nested-Arrays:-A-Novel:10}. It is obtained by nesting two uniform waveforms with different inter-element spacing. The inter-preamble spacing in the two-level NVW is
\begin{equation}
d \mathcal{M}_{\na} = \{ 1^{(M_1)}, (M_1+1)^{(M_2)}\},
\end{equation}
where $M_1$ is the number of preambles in the first uniform waveform and $M_2$ is the number of preambles in the second uniform waveform.
The VP count of the NVW is
\begin{equation}
\vert \mathcal{C}_{\na} \vert = M_2(M_1+1) -1.
\end{equation}
Based on VP count optimization, the optimal values for even $\vert \mathcal{M}_{\na} \vert$ is $M_1^* = M_2^* = \vert \mathcal{M}_{\na} \vert /2$ and for odd $\vert \mathcal{M}_{\na} \vert /2$ is $M_1 = M_2 +1 = (\vert \mathcal{M}_{\na} \vert  - 1)/2$.
The fraction of communication data symbols, $1 -\mu^*_{\na}$, for the NVW in a CPI is
\begin{equation}\label{eq:alphaII}
1- \mu_{\na}^* = 1 - \frac{( M_1^*+M_2^*) P \Ts+M_{\na}^*  T_\mathrm{IFS}}{T},
\end{equation}
where $M_{\na}^*$ is the optimal minimum $\vert \mathcal{M}_\rmV\vert$ for the NVW. In Section~\ref{sec:Simulation}, we will $M_1^*$ and $M_2^*$ values obtained using the VP count-based optimization and the CRB-based optimization.



Considering only the class of MUSIC-type algorithms for velocity estimation using non-uniform waveform design, we adopt 
the following velocity estimation methods: direct augmentation based MUSIC (DA-MUSIC) \cite{LiuVai:Remarks-on-the-Spatial-Smoothing:15}, spatial smoothing based MUSIC (SS-MUSIC) \cite{PalVai:Nested-Arrays:-A-Novel:10}, and direct-MUSIC \cite{VaiPal:Direct-MUSIC-on-sparse-arrays:12}. Since DA-MUSIC and SS-MUSIC share the same asymptotic first- and second- order error statistics \cite{WanNeh:Coarrays-MUSIC-and-the-Cram:17} and DA-MUSIC has reduced computational complexity, this paper only focuses on DA-MUSIC. The DA-MUSIC technique is applied on the virtual preambles in the co-waveform domain, whereas direct-MUSIC is applied directly on the physical non-uniformly spaced preambles. DA-MUSIC can be applied for $K\leq M_\rmV$ and $K > M_\rmV$, whereas direct-MUSIC is applicable only for $K \leq M_\rmV$ under certain conditions (namely that no two sources can be separated by multiples of $2\pi/Q$ where $Q$ is an integer which depends on the non-uniform preambles placement). Direct-MUSIC, however, is sometimes more accurate than DA-MUSIC for $K\leq M_\rmV$, as shown in Section~\ref{sec:Simulation}.  

\section{Simulation results}
\label{sec:Simulation}
%

In this section, we evaluate the performance of non-uniform virtual waveform design as compared to the uniform waveform design for different target and SNR scenarios. For illustration purposes, we consider simulation parameters based on the IEEE 802.11ad standard \cite{ieee2012wireless} in application to automotive scenarios \cite{PatTorWan:Automotive-Radars:-A-review:17}.  In particular, we consider a carrier frequency of 60 GHz, a sampling rate of 1.76 GHz, $K$ target velocities that are equally spaced between -45 m/s and 50 m/s, a radar cross-section of 10~dBsm, a communication receiver distance of 50 m and a radar target distance up to 100~m.  We use a coherent processing interval of 1 ms with $M$ varying between 3 and 40.

\subsection{Performance trade-off}
First, the system design trade-off between radar and communication MMSE performance metrics for different virtual waveform designs is studied for various target and SNR parameters. Then, the convex approximation of the design trade-off curve for improved JCR performance is described. Lastly, MMSE achievable by non-uniform designs are explored. In particular,  sparse ($K/M << 1$) and dense ($K/M \approx 1$ or $K/M >> 1$) target scenarios as well as low and high SNR use cases are investigated. 

\begin{figure}[!t]
\begin{minipage}[h]{\columnwidth}
\centering
\includegraphics[width=0.7\columnwidth]{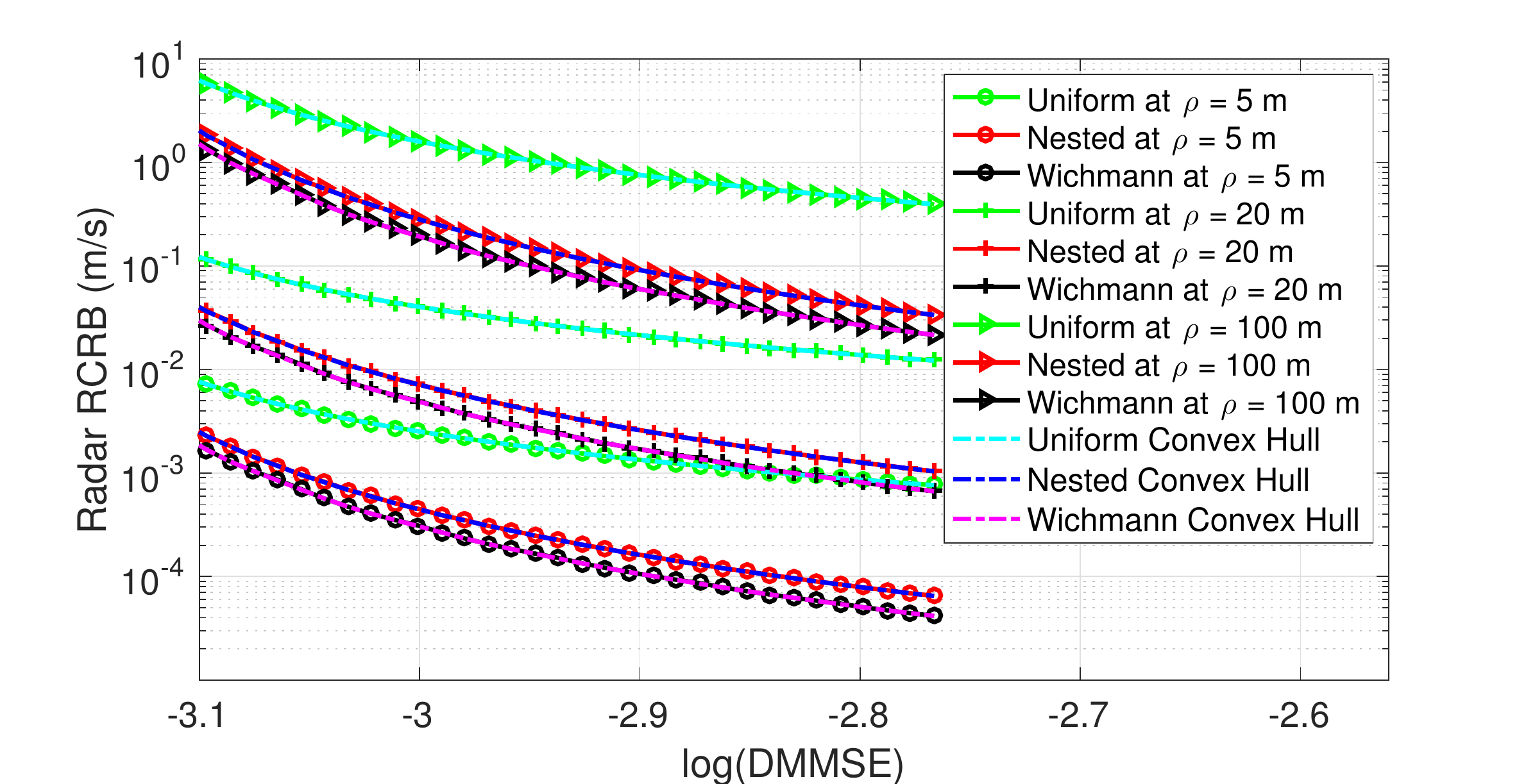}
\newline
 {(a) Single-target with $K=1$ }
   \end{minipage}
\begin{minipage}[h]{\columnwidth}
\centering
\includegraphics[width=0.7\columnwidth]{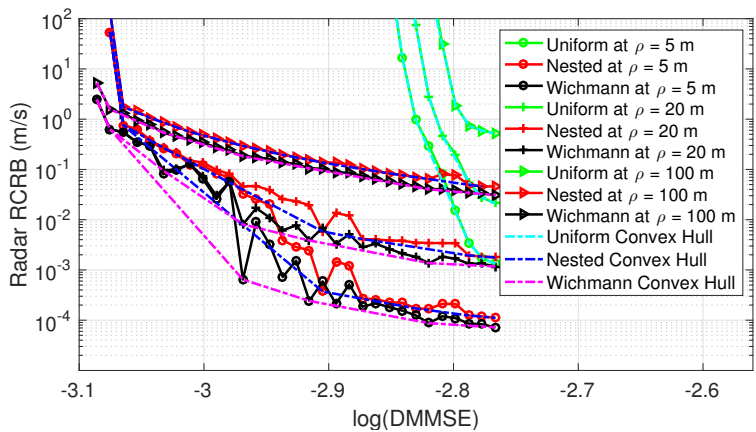}
\newline 
 {(b) Multi-target with $K=30$ } 
 \end{minipage}
   \caption{Trade-off curves between the radar RCRB and the communication DMMSE along with their corresponding convex hulls for different target counts and distances: (a) Trade-off curves are approximately convex for small target count $K$ for both uniform and non-uniform waveforms, (b) At large $K$, however, the non-convexity increases for non-uniform waveform. }
  \label{fig:Sim_TradeRadCom}
  \vspace{-.7em}
  \end{figure}
  
Fig.~\ref{fig:Sim_TradeRadCom} depicts the trade-off between the radar root CRB (RCRB) and the communication DMMSE metrics for uniform, nested, and Wichmann waveforms. In particular, radar target distances of 5~m, 20~m, and 100~m are considered for target count $K=1$ in Fig.~\ref{fig:Sim_TradeRadCom}(a) and for $K = 30$ in Fig.~\ref{fig:Sim_TradeRadCom}(b). For $M$ frames in a CPI, the optimal VP count-based nested waveform with parameters $M_1^*$ and $M_2^*$, as well as Wichmann waveform with parameters $p^*$ and $q^*$ are considered for simplicity in this example.  Later in Fig.~\ref{fig:Sim_Sol_DoF}, these parameters are also optimized based on the CRB and compared with that optimized based on the VP count of virtual waveforms. 

Fig.~\ref{fig:Sim_TradeRadCom} indicates that the trade-off between radar and communication is most relaxed in case of Wichmann virtual waveform, followed by nested virtual waveform and lastly uniform waveform. For a single target scenario in Fig.~\ref{fig:Sim_TradeRadCom}(a), we see that the advantage of virtual waveforms over uniform one is more significant as the communication DMMSE (higher $M$) worsens. In a multi-target scenario, the radar CRB for uniform waveform exists only for high communication DMMSE with $M>K$, whereas the radar CRBs for virtual waveforms exist even for low communication DMMSE with $M<K$. At low communication DMMSE, we also observe that the radar CRBs achieved by virtual waveforms saturate at high SNR and large $K/M$ ratio. 
 

Fig.~\ref{fig:Sim_TradeRadCom} also explores the convex hull of the trade-off curve between the radar RCRB and the communication DMMSE for different waveforms. Fig.~\ref{fig:Sim_TradeRadCom}(a) indicates that in a sparse target scenario with $K = 1$, the trade-off curves are approximately convex. In the case of a dense target scenario with $K=30$ in Fig.~\ref{fig:Sim_TradeRadCom}(b), however, the trade-off curves deviate farther from convexity as the radar target distance decreases. The trade-off curve for a uniform waveform is more visibly convex than that for the non-uniform waveforms.

The non-convexity in the trade-off curves is either because of the occurrence of non-decreasing RCRB points with increasing communication DMMSE or due to the saturation effect observed before the threshold point at small target distances and high $K/M$ ratio. Using the convex hull, the not so beneficial trade-off points ($M$) values, such as the non-decreasing CRB points or the points in the saturation region can be discarded. Additionally, the convex hull solution is achievable by using time-sharing or probabilistic occurrence techniques on the extreme points of the convex hull, similar to multi-user communication rate optimization \cite{brehmer2012utility}. Therefore, the convex hull approach will enable enhanced optimal solutions for JCR waveform designs by achieving a more relaxed trade-off curve between the radar RCRB and the communication DMMSE, as seen in Fig.~\ref{fig:Sim_TradeRadCom}.

\begin{figure}[!t]
\begin{minipage}[h]{0.5\columnwidth}
\centering
\includegraphics[width=0.7\columnwidth]{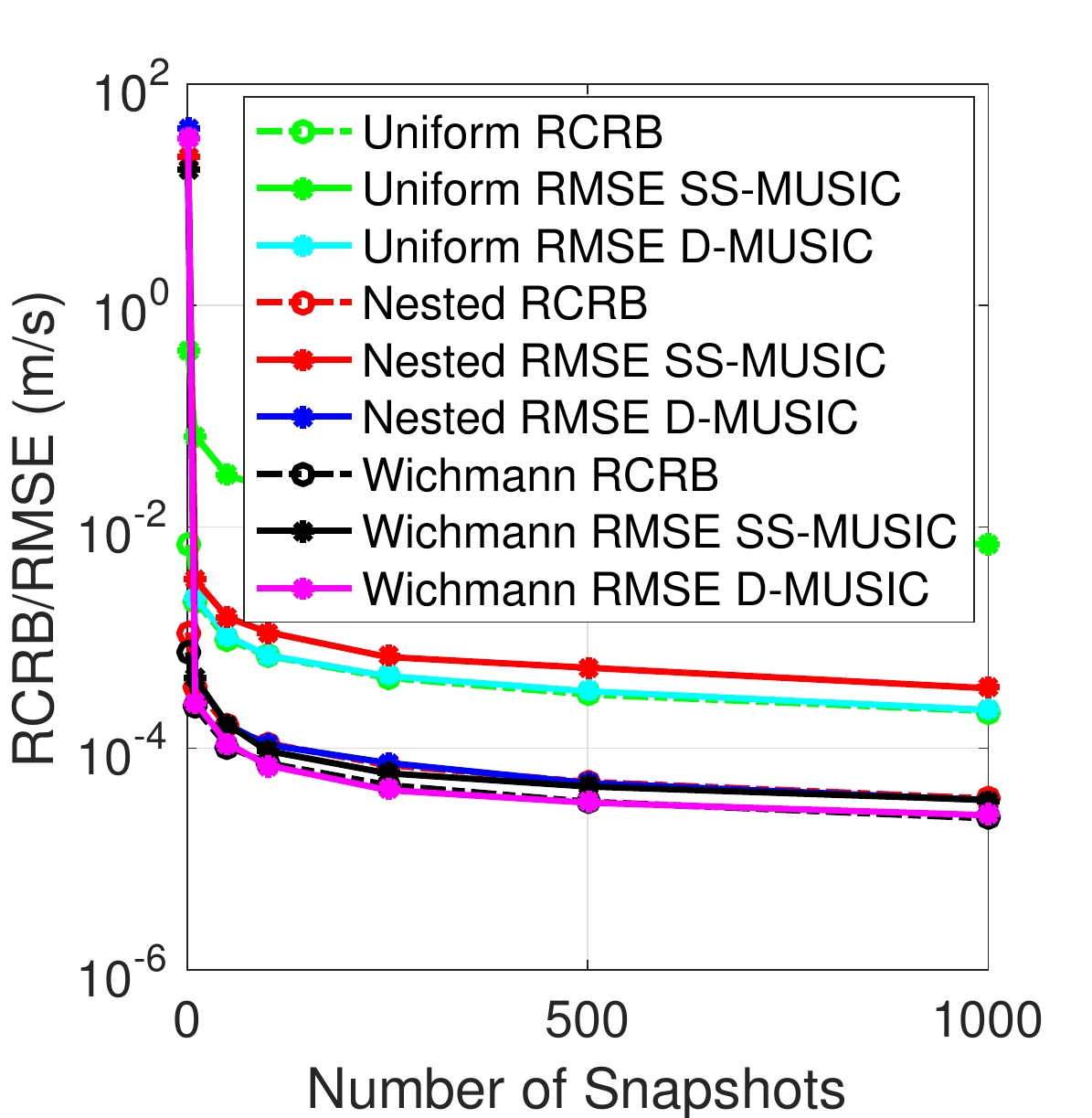}
\newline
 {(a) $K = 2$ and $\rho = 5$ m}
   \end{minipage}
\begin{minipage}[h]{0.5\columnwidth}
\centering
\includegraphics[width=0.7\columnwidth]{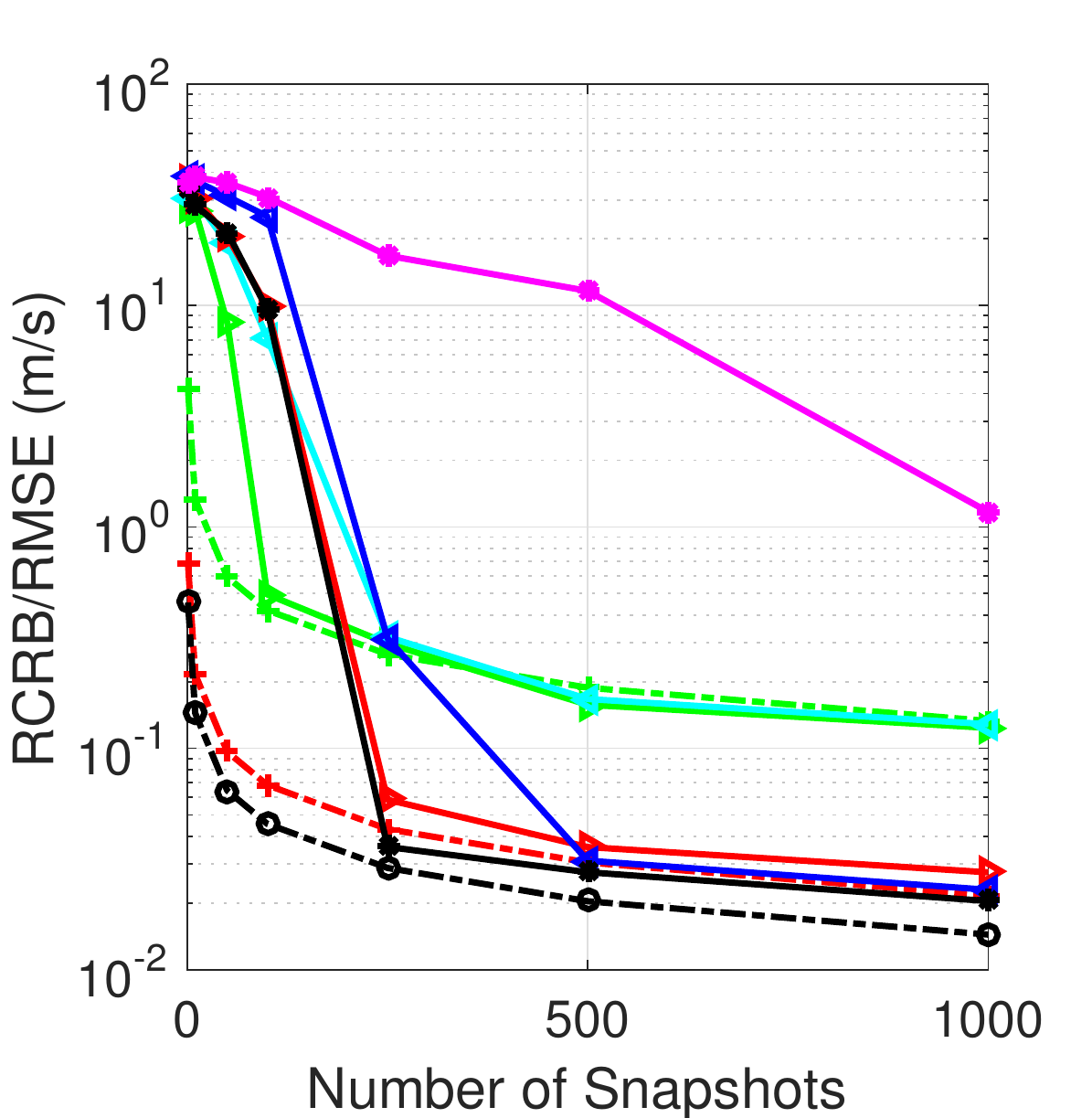}
\newline 
 {(b) $K = 2$ and $\rho = 100$ m} 
 \end{minipage}
   \caption{Comparison between RMSE of MUSIC-based algorithms with the RCRB for a two-target scenario at different target distances. The MUSIC-based algorithms with a small number of snapshots achieved the RCRB at high SNR, while they needed a higher number of snapshots at low SNR.}
  \label{fig:Sim_Algo}
  \vspace{-.8em}
  \end{figure}
 
Fig.~\ref{fig:Sim_Algo} compares the root mean square errors (RMSEs) achieved by MUSIC-based algorithms with the corresponding radar CRBs for all three tested waveforms in a two-target scenario with $M = 20$ and varying number of snapshots and target distances. In Fig.~\ref{fig:Sim_Algo}(a), Direct-MUSIC, in general, achieves RCRB more efficiently than DA-MUSIC at high SNR and is very close to the RCRB even with a small number of snapshots. In the low SNR scenario In Fig.~\ref{fig:Sim_Algo}(b), however, DA-MUSIC performs more efficiently than direct-MUSIC and it takes relatively more snapshots to approach the RCRB. The achievability of the RCRB for low SNR using MUSIC-based algorithms is more efficient in the case of uniform waveform as compared to virtual waveforms. 

The RMSE curves of MUSIC-based algorithms suggest that applicability of the CRB-based optimization for joint waveform design depends on the number of available snapshots. The MUSIC-based algorithms are used here for illustrating the achievability of the RCRB, but of course achievability of the optimal RCRB and communication DMMSE trade-off points can be enhanced by using other algorithms, such as orthogonal matching pursuit and nuclear norm minimization, and also by exploiting waveforms with holes, such as Golomb and coprime waveforms. This is, however, out-of-scope of this paper and is left for future work.


\subsection{Optimal waveform designs}
In this subsection, we investigate the optimal solutions for three MMSE-based waveform design formulations proposed in Section~\ref{sec:Problem}. We also compare our MMSE-based waveform optimization formulation with the more traditional VP count-based formulation.

\subsubsection{Weighted average optimization-based design}

In this example, we explore the optimal communication DMMSE and radar CRB via weighted average problem formulation \eqref{eq:weighted}. First, we explore the effect of weighting on the optimal solutions, followed by the effect of target count and SNR on the optimal solutions for all three tested waveforms.

\begin{figure}[!t]
\begin{minipage}[h]{\columnwidth}
\centering
\includegraphics[width=0.7\columnwidth]{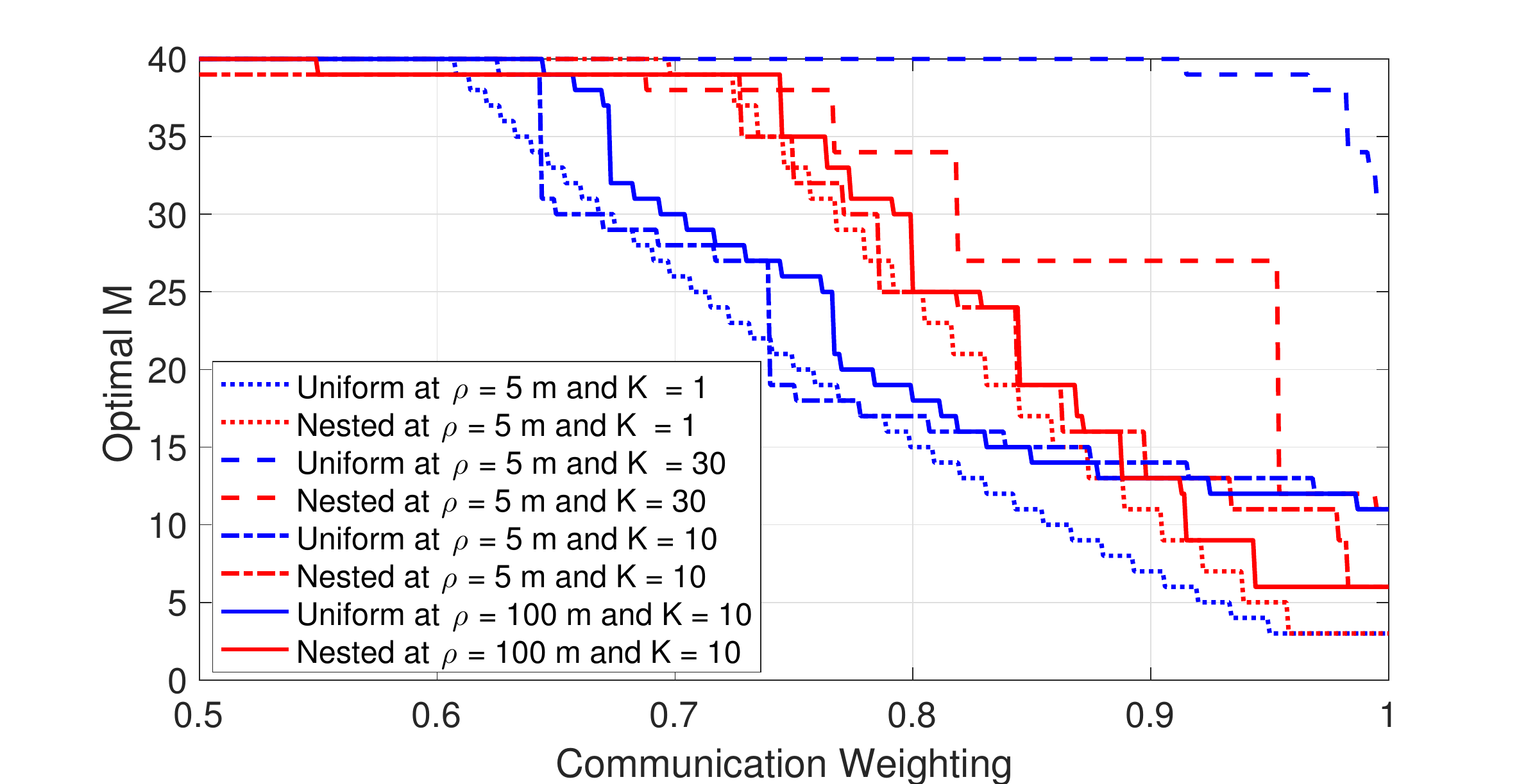}
\newline
 {(a) Optimal $M$}
   \end{minipage}
\begin{minipage}[h]{\columnwidth}
\centering
\includegraphics[width=0.7\columnwidth]{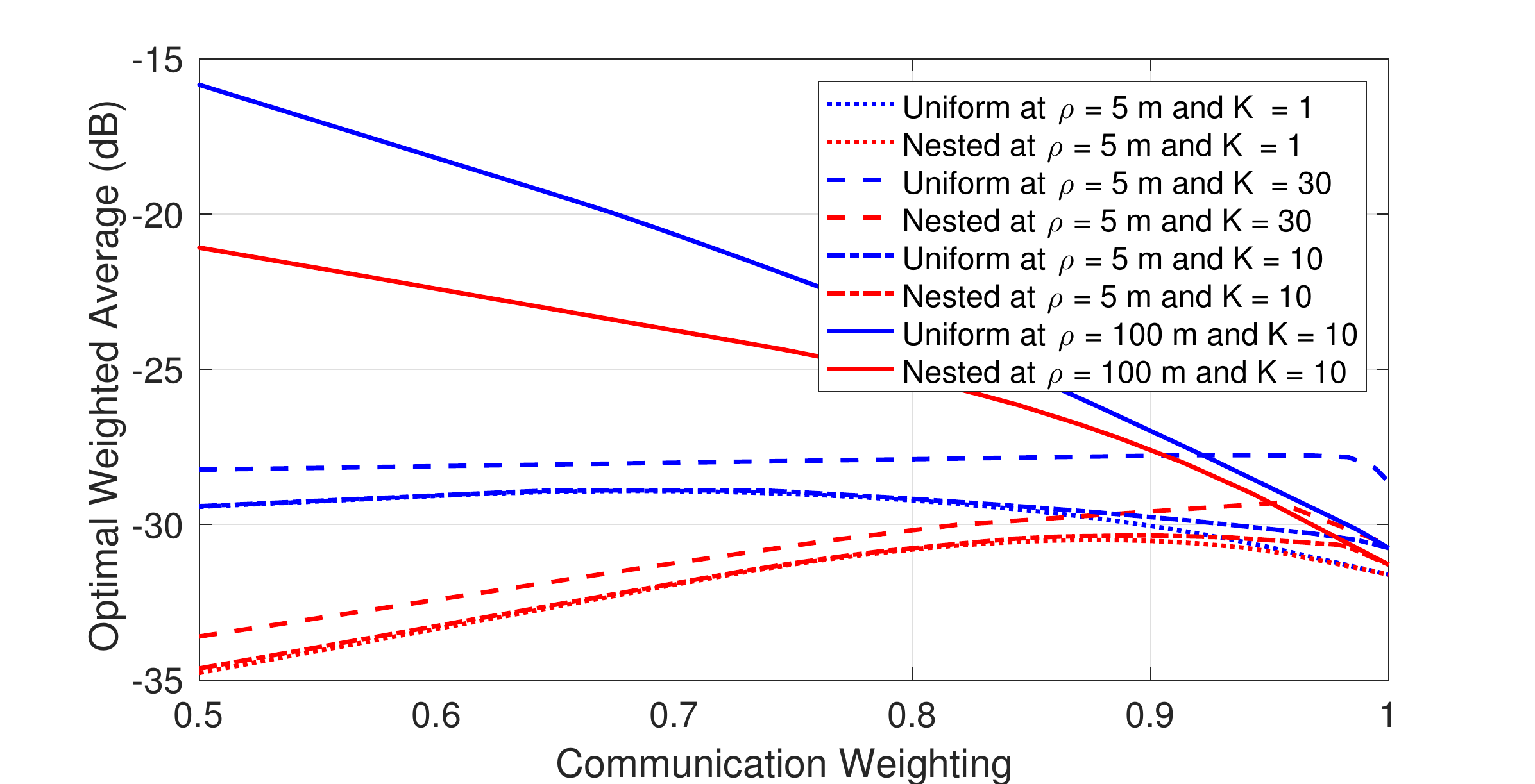}
\newline 
 {(b) Optimal weighted average} 
 \end{minipage}
   \caption{Optimal number of frames $M$ and weighted average for uniform and nested waveforms with different normalized communication weightings: (a) The optimal $M$ decreases with varying step-size as the communication weighting increases, (b) The advantage of the nested waveform over the uniform one decreases as the communication weighting approaches to 1.}
  \label{fig:Opt_Sol3_w}
  \vspace{-.7em}
  \end{figure}
Figs.~\ref{fig:Opt_Sol3_w}(a) and \ref{fig:Opt_Sol3_w}(b) show the optimal number of frames $M$ for uniform and nested waveforms with different normalized communication weightings $0.5 \leq \omega_{\comm} \leq 1$ at target counts $K= \{1,10, 30 \}$ for target distance $\rho = 5$~m, as well as at target distances $\rho = \{5, 100\}$ meters for $K = 10$. The optimal $M$ for different waveforms decreases from highest possible $M = 40$ used in a CPI for $\omega_{\comm} = 0$ to the lowest feasible $M$ that satisfies the CRB existence condition for $\omega_{\comm} = 1$. The optimal $M$ varies significantly with $K$ because of the CRB existence and the saturation effect observed at high SNR and high $K/M$ ratio. The rate of decrease in the optimal $M$ in a sparse target scenario is smaller for uniform waveform as compared to non-uniform waveforms because of the steeper decrease in the radar RCRB with increase in communication DMMSE achieved by non-uniform waveforms. The step size of the decrease in the optimal $M$ with increasing communication weighting for a dense target scenario at high radar SNR is generally small for uniform waveform as compared to non-uniform waveforms due to the approximate convexity of the weighted average JCR performance metric for uniform waveform. The step-size also depends on the presence of non-decreasing radar CRB points with increasing communication DMMSE in the corresponding trade-off curve.

\begin{figure}[!t]
\begin{minipage}[h]{0.5\columnwidth}
\centering
\includegraphics[width=0.7\columnwidth]{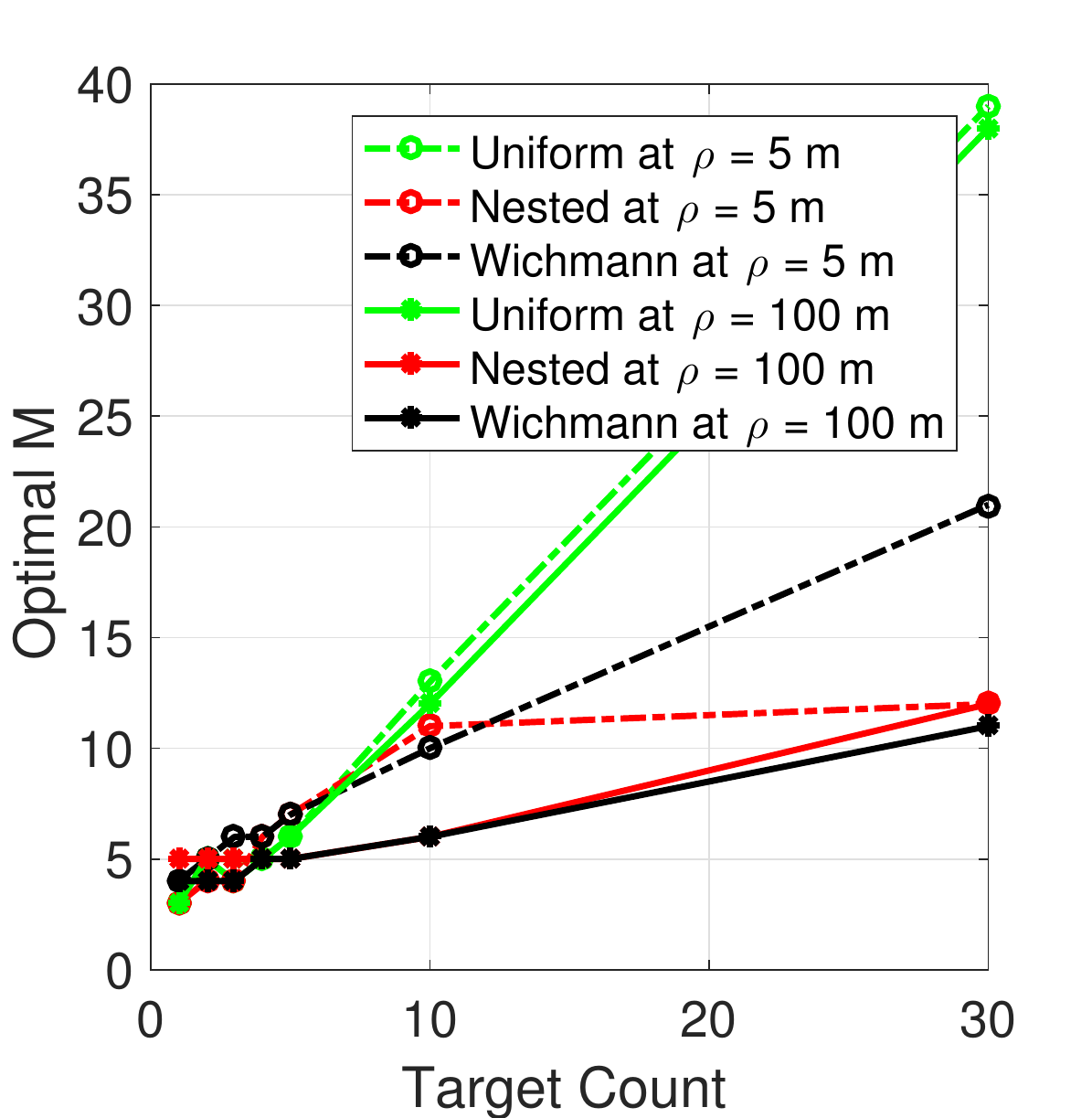}
\newline
 {(a) Variation with $K$}
   \end{minipage}
\begin{minipage}[h]{0.5\columnwidth}
\centering
\includegraphics[width=0.7\columnwidth]{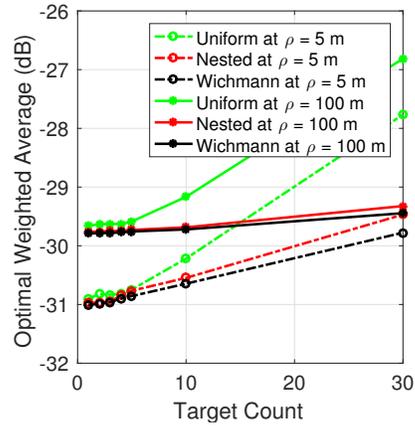}
\newline 
 {(b) Variation with $K$} 
 \end{minipage}
  \vfill

\begin{minipage}[h]{0.5\columnwidth}
\centering
\includegraphics[width=0.7\columnwidth]{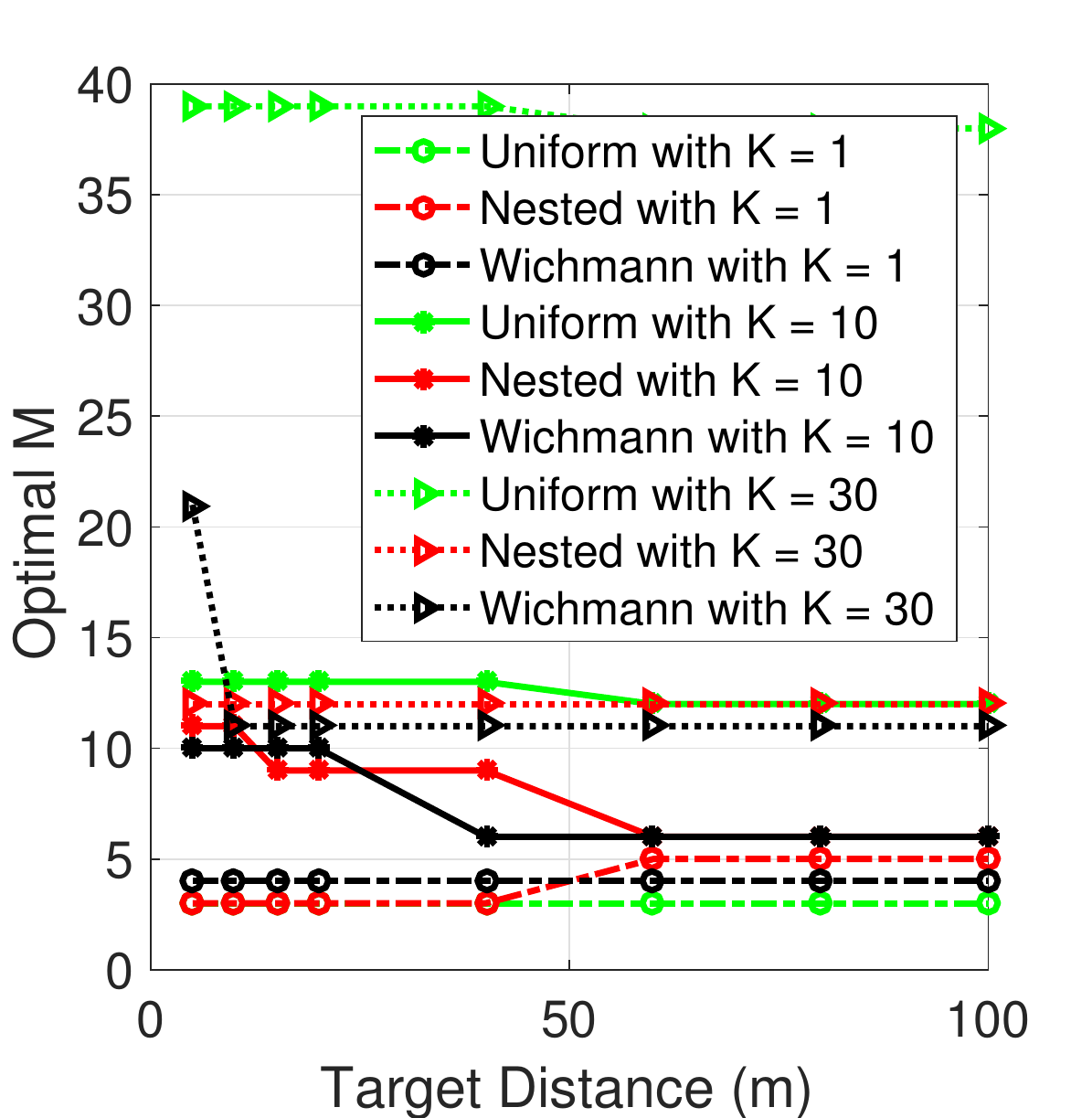}
\newline
 {(c) Variation with $\rho$}
   \end{minipage}
\begin{minipage}[h]{0.5\columnwidth}
\centering
\includegraphics[width=0.7\columnwidth]{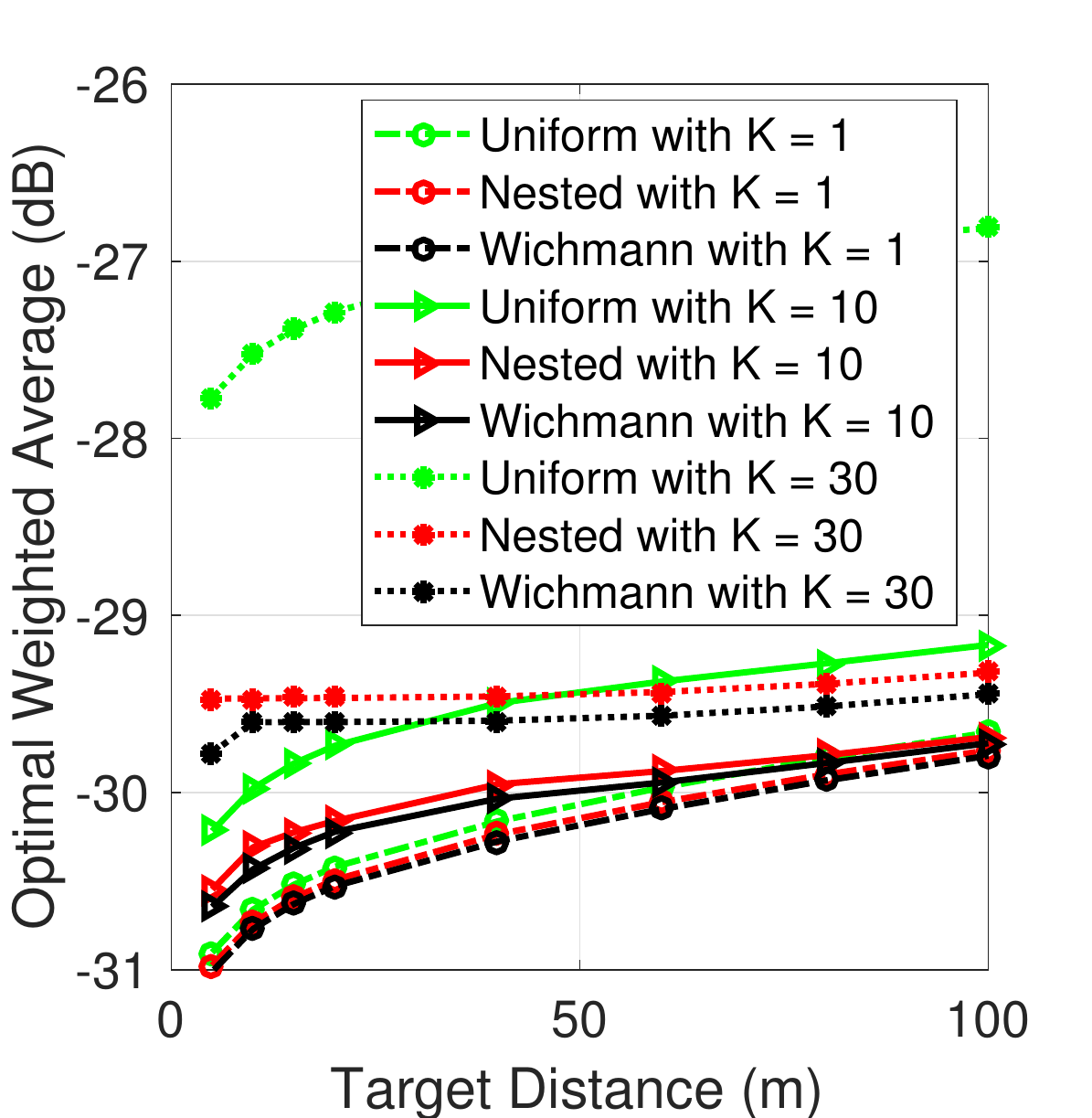}
\newline 
 {(d) Variation with $\rho$} 
 \end{minipage}
   \caption{Optimal $M$ and corresponding weighted average JCR MMSE with communication weighting of 0.96. The advantage of non-uniform waveform over uniform one increases with the target count, whereas it reduces with radar SNR at high target count due to the saturation effect.}
  \label{fig:Sim_Sol3_Dist_K}
  \vspace{-.7em}
  \end{figure}

  
Fig.~\ref{fig:Opt_Sol3_w}(b) shows the optimal weighted average of the communication DMMSE and the radar RCRB for different normalized communication weightings $0.5 \leq \omega_{\comm} \leq 1$ at $K = 1$, 10, and 30 for $\rho = 5$~m, as well as at $\rho = 5$~m and $100$~m for $K = 10$. The advantage of nested waveform over uniform at a given $K$ and $\rho$ decreases as communication weight $\omega_\mcom$ approaches to 1. The performance of all waveforms tested converge exactly for $\omega_\mcom = 1$ at $K=1$, because the lowest possible $M$ is used in this case. At $\omega_\mcom = 1$, the gap between the optimal performance of uniform and non-uniform waveforms increases with $K$ due to higher $M$ needed for the CRB to exist, while it remains constant with $\rho$ at $K=10$. For most of the scenarios, nested waveform performs the best and uniform one performs the worst. The insights derived for nested waveform similarly can be extended for Wichmann waveform, and Wichmann waveform generally performs better than nested as can be seen in Figs.~\ref{fig:Sim_Sol3_Dist_K}(a)-(d).

Figs.~\ref{fig:Sim_Sol3_Dist_K}(a) and \ref{fig:Sim_Sol3_Dist_K}(b) show the variation of optimal $M$ and corresponding weighted average JCR MMSE-based metrics with $K$ for communication weighting of 0.96. The optimal $M$ increases with target count in most of the cases, except for the nested waveform at high SNR and large $K$ that suffers from saturation effect. The choice of optimal $M$ in the case of Wichmann waveform as compared to uniform one increases with target count, and the optimal $M$ for all three waveforms tested converges to the same value at low target count. The optimal weighted average for uniform waveform is the worst (largest) in all the cases. The optimal weighted average degrades with increasing target count, and the advantage of non-uniform waveform over uniform one increases with $K$, whereas it reduces with radar SNR at high $K$ due to the saturation effect.

Figs.~\ref{fig:Sim_Sol3_Dist_K}(c) and \ref{fig:Sim_Sol3_Dist_K}(d) depict the variation of optimal $M$ and corresponding weighted average JCR MMSE-based metrics with $\rho$ for communication weighting of 0.96. Fig.~\ref{fig:Sim_Sol3_Dist_K}(c) shows that the optimal $M$ increases with distance at low target count and decreases with distance at high target count. This effect is due to the radar RCRB degradation that happens because of pathloss increasing at large distances for low target count and saturation effect increasing at small distances for high target count. Fig.~\ref{fig:Sim_Sol3_Dist_K}(d) demonstrates that the optimal weighted average for uniform waveform is the worst in all the cases. The optimal weighted average for all three waveforms tested generally improves with decreasing distance. For $K=30$, however, the saturation effect can be seen for nested and Wichmann waveforms. The rate of improvement in optimal weighted average with increasing radar SNR reduces with growing $K$ for non-uniform waveforms, whereas it remains constant for uniform waveform.

\subsubsection{Radar CRB constrained optimization-based design}
  \begin{figure}[!t]
\begin{minipage}[h]{\columnwidth}
\centering
\includegraphics[width=0.7\columnwidth]{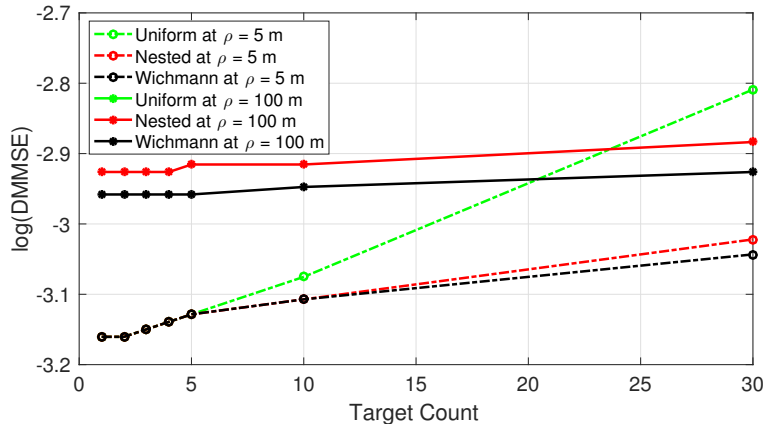}
\newline
 {(a) Variation with target count}
   \end{minipage}
\begin{minipage}[h]{\columnwidth}
\centering
\includegraphics[width=0.7\columnwidth]{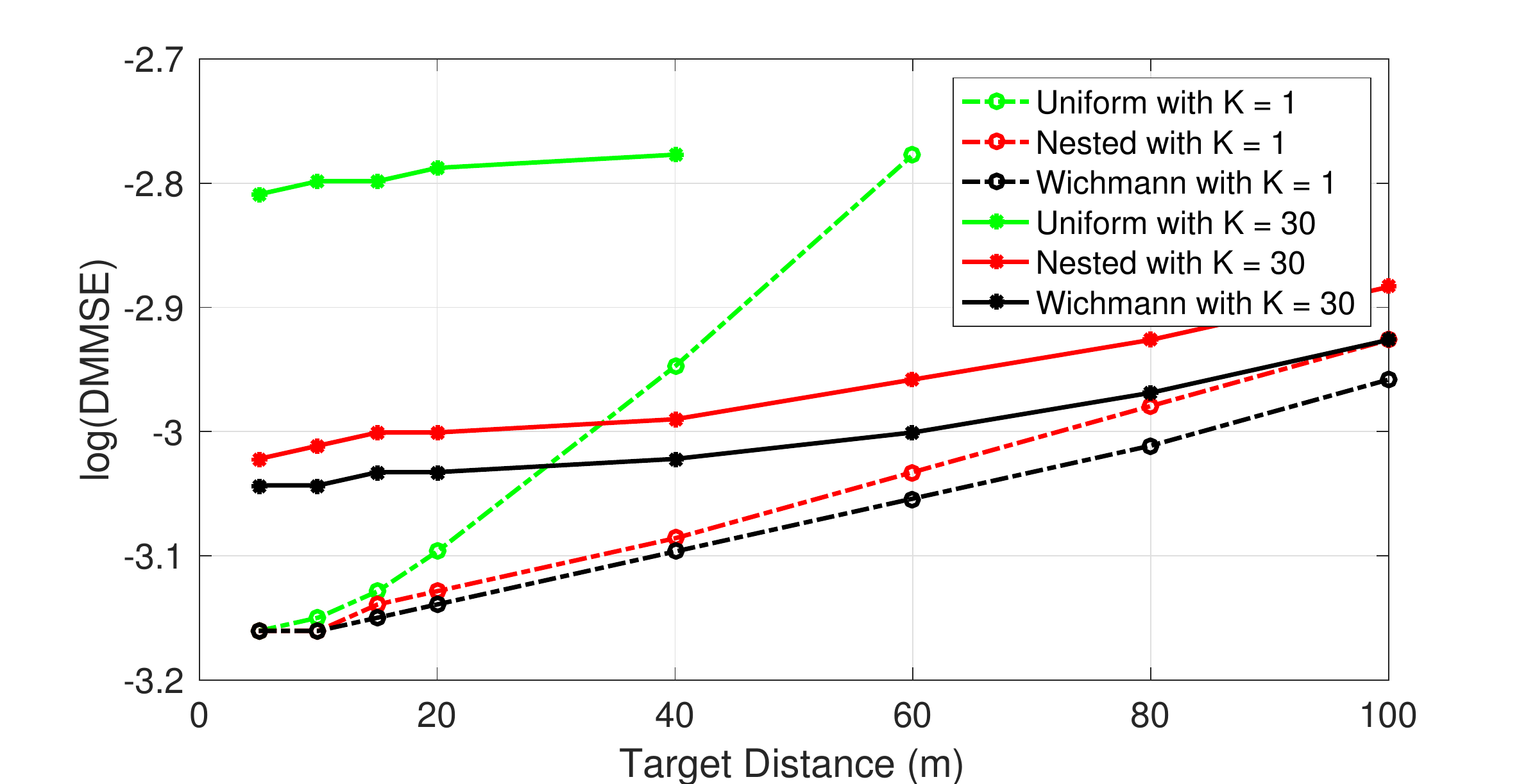}
\newline 
 {(b) Variation with target distance} 
 \end{minipage}
   \caption{Optimal communication DMMSE (and corresponding optimal $M$) for a given minimum radar CRB constraint $\Upsilon_\mrad = 1.5$~cm$^2$/s$^2$. The advantage of virtual waveforms over uniform one increases with target count at small target distance and also with distance at low target count.}
  \label{fig:Sim_Sol3_K_Dist}
  \vspace{-.7em}
  \end{figure}
In this example, we investigate optimal communication DMMSE (and corresponding optimal $M$) for a given minimum radar CRB constrained problem formulation \eqref{eq:RadConst} in different target density and SNR scenarios. Fig.~\ref{fig:Sim_Sol3_K_Dist} shows the optimal DMMSE solution for a minimum radar CRB of -18~dB. Wichmann waveform performs the best, followed by nested waveform. The advantage of virtual waveforms over uniform one increases with target count at high SNR and also with distance at low target density. Performances of all three waveforms tested (Wichmann, nested, and uniform) converge at low target count and small target distance. The optimal DMMSE increases with growing target count and decreasing SNR. Optimal communication DMMSE, however, is less effected by target count at small distances, as compared to the large distances.  Additionally, the rate of increase of optimal communication DMMSE with increasing target distance is faster at lower $K$. These effects can be explained using the saturation effect at high $K/M$ ratio and high SNR. The effect of the preamble count can also be seen on the feasibility of uniform solutions. The insights for optimal $M$ are similar due to its linear relation with communication DMMSE. 

\subsubsection{Communication DMMSE constrained optimization-based design}
In this example, we examine the optimal radar RCRB achieved by all three waveforms tested for a given minimum communication DMMSE constrained problem formulation \eqref{eq:CommConst}. First, we study the effect of different DMMSE constraints and target count on the optimal radar RCRB at a given target distance. Then, we explore the effect of varying target count and distance on the optimal radar RCRB for a given DMMSE constraint. 

 \begin{figure}[!t]
 \begin{minipage}[h]{\columnwidth}
\centering
\includegraphics[width=0.7\columnwidth]{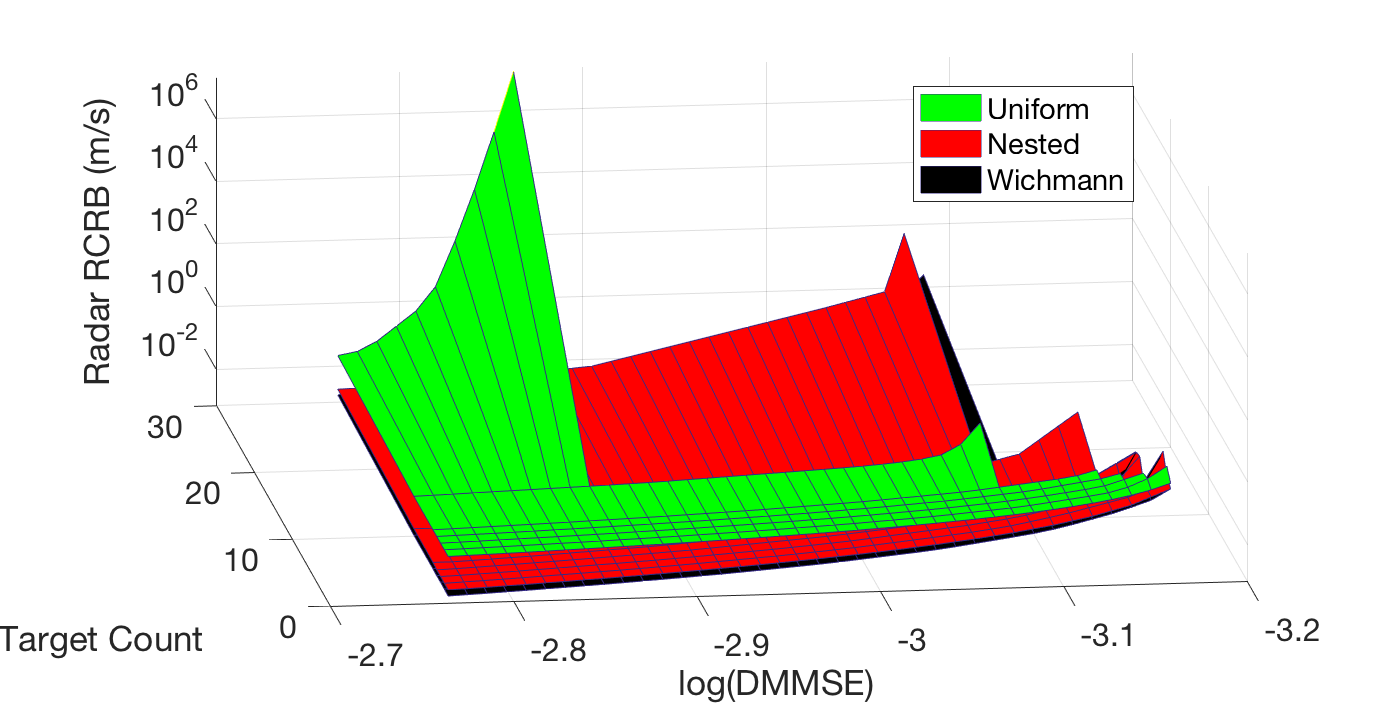}
\newline
 {Variation with $K$ and DMMSE for $\rho = 20$~m. }
 \end{minipage}
   \caption{The mesh plot of the optimal radar RCRB for various minimum communication DMMSE constraints at different $K$ scenarios for a radar target distance of 20~m. The gap between the performances of all the three tested waveforms improve with increasing target count and communication DMMSE.}
  \label{fig:Sim_Sol2_Mesh}
  \vspace{-.7em}
  \end{figure}

 Fig.~\ref{fig:Sim_Sol2_Mesh} depicts the optimal radar RCRB variation with the target count $K$ and communication DMMSE constraint $\Upsilon_\mcom$ at a radar target distance $\rho = 20$~m. The feasibility of the optimal radar RCRB solution depends on the target count. For $2K\leq \vert \mathcal{C}_\rmV \vert$, where $\vert \mathcal{C}_\rmV \vert$ is the VP count of the given waveform design, the optimal $M$ that minimizes the RCRB for a given communication DMMSE may exist and it corresponds to the maximum $M$ that satisfies $\Upsilon_\mcom$ constraint on the convex hull of the JCR trade-off curve for a given CPI.  Therefore, the feasibility of the Wichmann waveform is the highest, whereas it is lowest for the uniform one. Additionally, the Wichmann waveform achieves the optimal RCRB better than the nested one, followed by the uniform one. The gap between the performance of all the three tested waveforms improve with increasing target count and communication DMMSE. Additionally, the performance of all three waveforms tested converge at low communication DMMSE.

  \begin{figure}[!t]

\begin{minipage}[h]{\columnwidth}
\centering
\includegraphics[width=0.7\columnwidth]{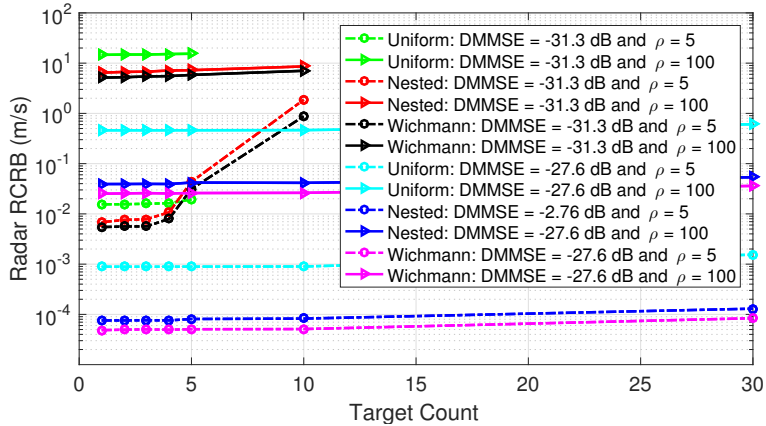}
\newline
 {(a) Variation with $K$ for $\rho =\{ 5, 100 \}$ m.}
   \end{minipage}
\begin{minipage}[h]{\columnwidth}
\centering
\includegraphics[width=0.7\columnwidth]{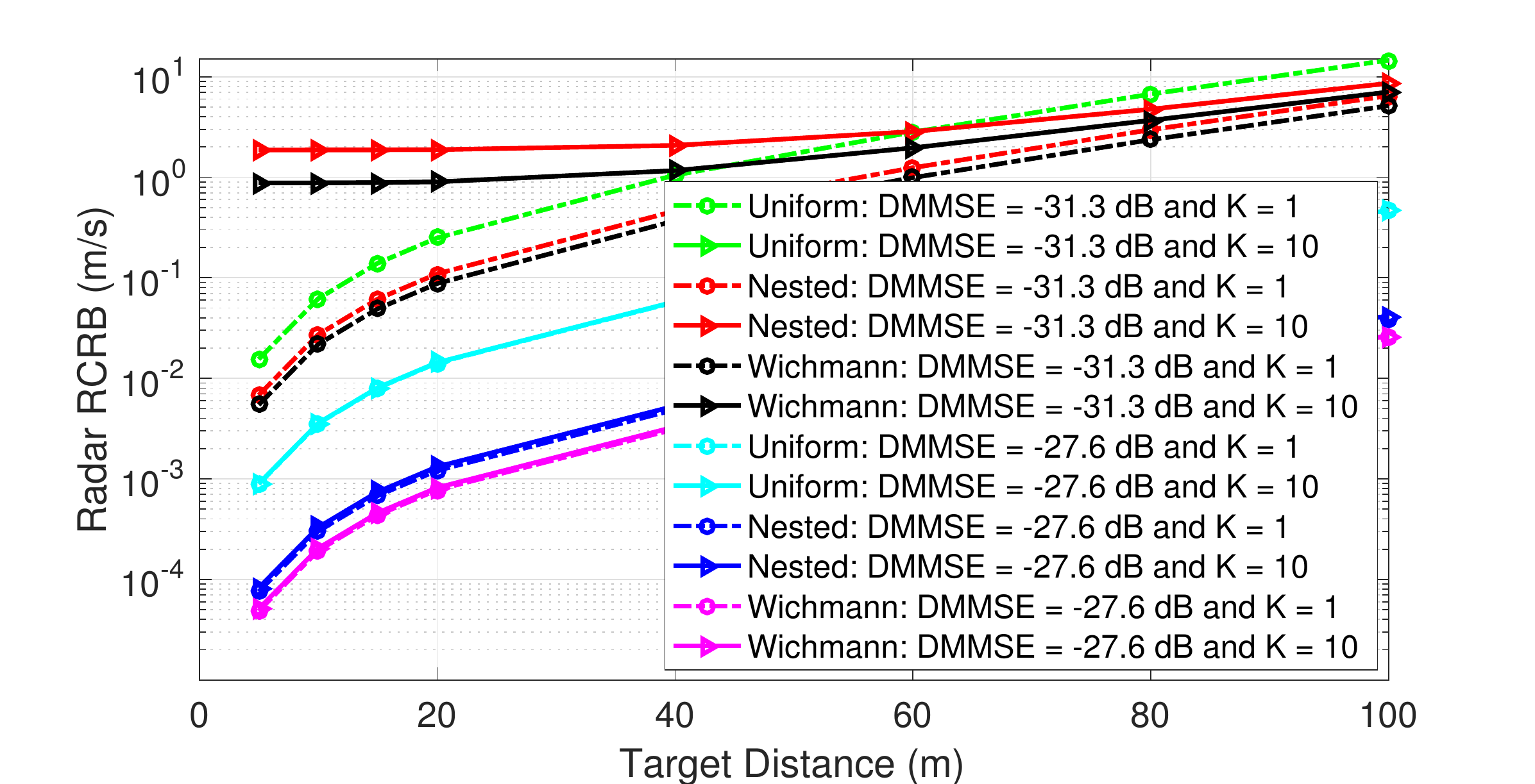}
\newline 
 {(b) Variation with $\rho$ for $K =\{ 1, 10 \}$.} 
 \end{minipage}
   \caption{Optimal radar RCRB for a communication DMMSE of -31.3~dB and -27.6~dB. The advantage of virtual waveforms over the uniform one is more than 10 dB at high communication DMMSE. }
  \label{fig:Sim_Sol2_K_Dist}
  \vspace{-.7em}
  \end{figure}
  
Fig.~\ref{fig:Sim_Sol2_K_Dist} show the optimal radar RCRB for all three tested waveforms with communication DMMSE constraint $\Upsilon_\mcom = -31.3$~dB and $\Upsilon_\mcom = -27.6$~dB at target counts $K= 1$ and $K = 30$ and radar target distances $\rho = 5$~m and 100~m. For DMMSE of -31.3~dB, the optimal $M = 6$, while for DMMSE of $-27.6$~dB, the optimal $M = 40$ for most of the scenarios. For target count $K = 30$ at radar distances $80$~m and $100$~m, however, the optimal $M = 39$ for nested waveform at DMMSE of $-27.6$~dB. This is because of the non-decreasing radar RCRB point at $M=40$ for $K = 30$. 

For all three waveforms tested, Fig.~\ref{fig:Sim_Sol2_K_Dist}(a) depicts that the optimal radar RCRB degrades with increasing target count and Fig.~\ref{fig:Sim_Sol2_K_Dist}(b) shows the optimal radar RCRB grows with increasing radar distance. In most cases, the Wichmann waveform performs the best and uniform one performs the worst. At high communication DMMSE with the optimal $M=40$, the Wichmann waveform achieved more than 10 dB improvement in the velocity estimation RCRB as compared to the uniform one. Fig.~\ref{fig:Sim_Sol2_K_Dist}(a) shows that the performances of all three waveforms tested converge at low communication DMMSE and high SNR ($\rho$ = 5~m) with high target density $K/M$ ($K$ = 5 and $M$ = 6), if the CRB exist for all three tested waveforms. Fig.~\ref{fig:Sim_Sol2_K_Dist}(b) shows the saturation effect at high SNR (small $\rho$) with high $K/M$ ($K = 10$ and $M = 6$) ratio for nested and Wichmann waveforms, while the uniform waveform is not feasible due to $K \geq M$.  The radar RCRB achieved for virtual waveforms at low communication DMMSE with the optimal $M = 6$ is effected by the change in target count $K$ for small target distances, whereas the optimal radar RCRB is only slightly effected at long distances.  At high communication DMMSE, however, this saturation effect is not observed.



\subsubsection{Comparison with VP count optimization-based design}
    \begin{figure}[!t]
\begin{minipage}[h]{\columnwidth}
\centering
\includegraphics[width=0.7\columnwidth]{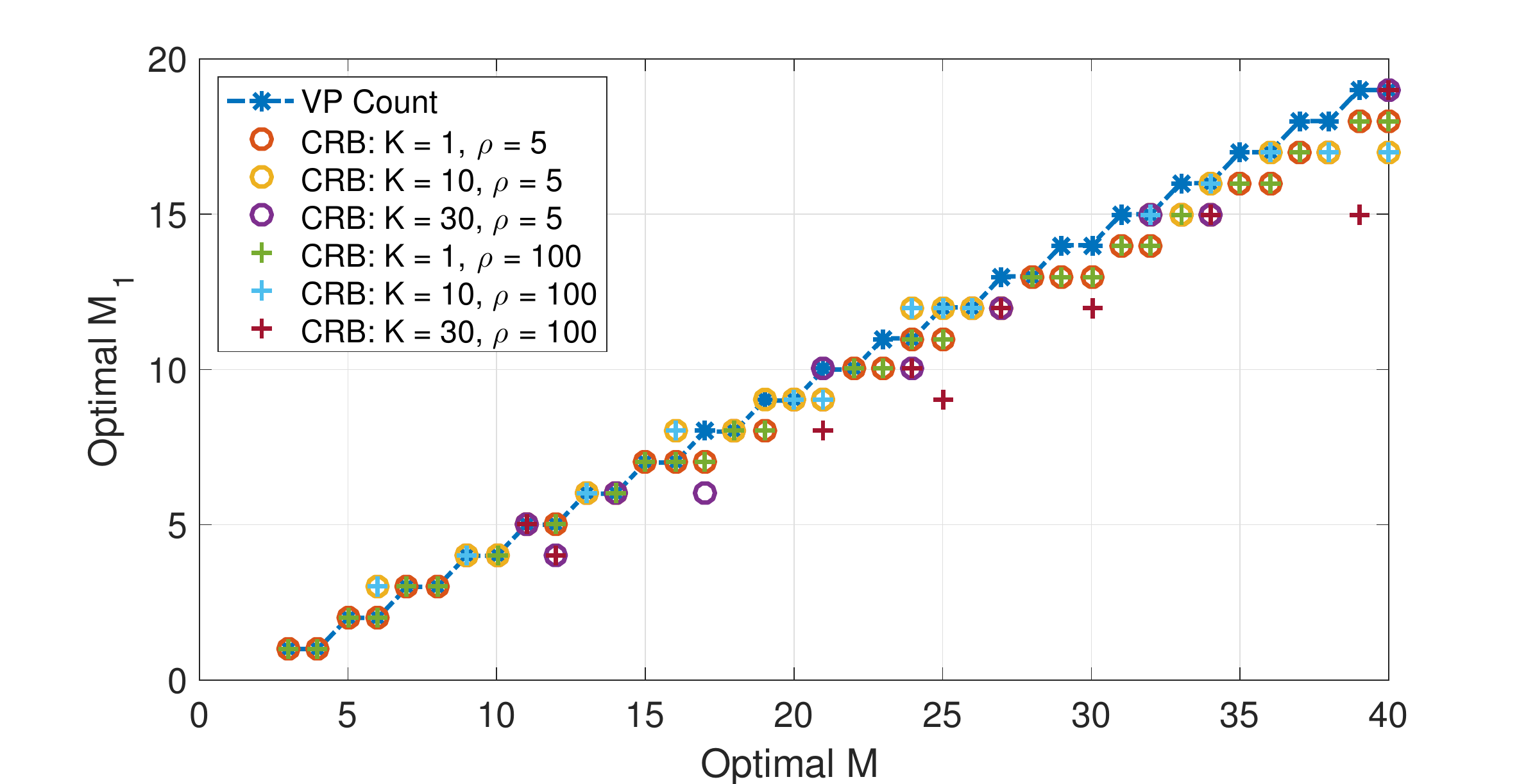}
\newline
 {(a) Optimal configuration parameter $M_1$}
   \end{minipage}
 \vfill
\begin{minipage}[h]{\columnwidth}
\centering
\includegraphics[width=0.7\columnwidth]{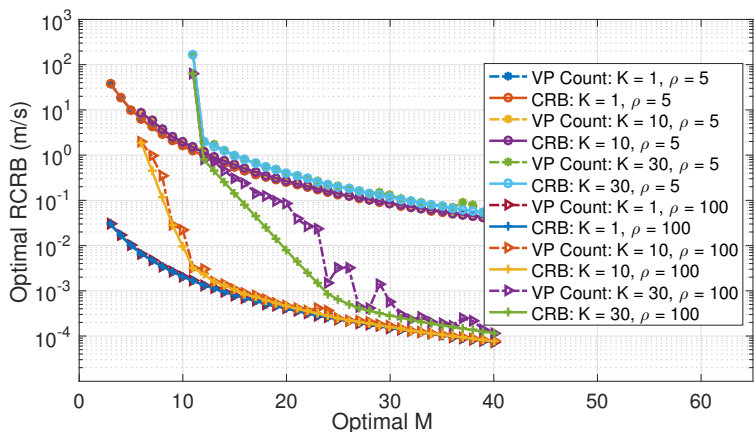}
\newline 
 {(b) Optimal radar RCRB} 
 \end{minipage}
   \caption{Comparison between the VP count-based optimization and the CRB-based optimization for nested waveform. The gap between the optimal solutions of both the optimizations grows with increasing target count and decreasing target distance.}
  \label{fig:Sim_Sol_DoF}
  \vspace{-.7em}
  \end{figure}
Fig.~\ref{fig:Sim_Sol_DoF} explores the optimal configuration parameter $M_1$ and radar RCRB that maximizes the VP count (or, equivalently degrees of freedom (DoF)) using nested waveform for a given $M$. The figures also compares the virtual preamble count-based optimization with its respective CRB-based communication DMMSE constrained optimization for different $K$ and radar SNR.  For odd $M$, the unique optimal $M_1$ solution using the VP count-based optimization is $(M-1)/2$ and for even $M$ there are two solutions $M/2$ and $M-1/2$. Fig.~\ref{fig:Sim_Sol_DoF} uses the smallest optimal $M_1$ as VP count optimization-based one and we see that this optimal $M_1$ increases step-wise linearly with $M$. For $K=1$, the optimal $M_1$ for both the optimizations match at small $M$ and start deviating a little for higher $M$. Fig.~\ref{fig:Sim_Sol_DoF}(a) demonstrates that for $K =1$, the radar RCRB for both problem formulations are very close to each other at high SNR. The solutions, however, start to deviate for high target count and high radar SNR. Similar insights can be drawn for Wichmann waveform with configuration parameters $p$ and $q$, but this study is excluded from the paper due to space constraint. Fig.~\ref{fig:Sim_Sol_DoF} suggests that the VP count optimization-based design can be used as a coarse estimate for the CRB-based communication DMMSE constrained solution.


\section{Conclusion and discussion}
In this paper, we proposed an adaptive virtual waveform design for a millimeter-wave joint communication-radar system that enjoys the benefit of a fully-digital baseband processing in the time-domain and a high available bandwidth. Our proposed waveform exploits only a few non-uniform preambles in a CPI and sparse sensing techniques to achieve high velocity estimation accuracy without reducing communication data rate much. We developed a novel communication DMMSE metric to accurately quantify the trade-off with radar CRB for a JCR waveform design. 

The performance trade-off curve between the radar CRB and communication DMMSE contained some non-convex points due to the occurrence of non-decreasing CRB points with increasing communication DMMSE or due to the radar CRB saturation at high SNR. To improve the optimal JCR performance, we discarded these undesirable non-convex points by using a convex hull approximation of the trade-off curve. Then, we formulated three different MMSE-based problems to optimize the trade-off between communication and radar: a minimum communication DMMSE constrained formulation, a minimum radar CRB constrained formulation, as well as a weighted MMSE average formulation. To reduce the computation complexity for finding optimal waveform solutions, we used specific waveform configurations -- the uniform waveform, the nested virtual waveform, and the Wichmann virtual waveform. Numerical results demonstrated that, in most cases, non-uniform waveforms perform much better than uniform waveforms, especially at low SNR and high target density. Additionally, we observed that the traditional virtual preamble count-based solution can be used as a coarse estimate of the optimal solution for our MMSE-based optimization problems. 

The results in this paper can be taken into account to design an adaptive virtual preamble that achieves simultaneous high communication data rate and super-resolution radar estimation for next-generation mmWave devices. For future work, the proposed framework can be extended to other virtual waveforms, such as the Golomb waveform and the coprime waveform. The achievability of their CRBs can be investigated using more advanced estimation algorithms, such as nuclear norm minimization. This may lead to better performance at a low number of snapshots. This work can also be extended for a more general Ricean fading mmWave channel model with a small number of scattering clusters. It would be interesting to see how the Ricean fading factor and the additional block sparse structure will impact the advantage of the virtual waveforms that exploit compressive sensing techniques on the channel covariance matrix.


\bibliographystyle{IEEEtran}
\bibliography{IEEEabrv,Heathabrv,refv9}
\end{document}